\documentclass[aps,prb,preprint]{revtex4-2}
\usepackage{graphicx} 

\usepackage{float}
\usepackage{amsmath}
\usepackage{amssymb}
\usepackage{bm}
\usepackage{epsfig}
\usepackage{graphicx}
\usepackage{color}
\usepackage[normalem]{ulem}

\usepackage[utf8]{inputenc}
\usepackage{enumitem}

\usepackage{color}

\def\be{\begin{equation}}
\def\ee{\end{equation}}
\def\bea{\begin{eqnarray}}
\def\eea{\end{eqnarray}}

\begin{document}
\def\affiIOFFE{Ioffe Institute, Russian Academy of Sciences, 194021 St.~Petersburg, Russia}

\title{ Exciton fine structure in nanocrystals: \\ 
effect of cuboidal and spheroidal shapes}
\author{M.~A.~Semina, O.~O.~Druzhinina, A.~A.~Golovatentko, A.~V.~Rodina}
\affiliation{\affiIOFFE}
\date{\today}

\begin{abstract}

We present the theory of the band-edge exciton fine structure in nanocrystals (NCs) with spheroidal and cuboidal shapes. The effects of the cubic symmetry of the crystal lattice, the cubic shape of the NC, and NC uniaxial anisotropy on the hole energy states and electron-hole exchange interactions are considered non-perturbatively. Symmetry analysis yields an effective Hamiltonian for the exciton fine structure, parameterized by one constant for hole energy splitting and five independent constants for   exchange interaction. Numerical calculations reveal that in uniaxially anisotropic zinc-blende NCs,  the sign of the  hole ground state splitting  depends on the material parameters and on the orientation of the anisotropy axis relative to the crystallographic axes. Beyond the conventional bulk cubically-symmetric contribution to the  exchange interaction, which originates from Bloch-function symmetry and is typically negligible, in nanocrystals, we identify the contribution arising from the cubic symmetry of the envelope wavefunction. This cubically symmetric envelope-induced short-range exchange is non-negligible in cuboidal NCs and induces a pronounced splitting of the dark exciton states.
We further analyze the influence of uniaxial anisotropy of the exchange constants on both the exciton fine structure and the oscillator strength.   Special attention is paid to NCs, where the anisotropy of the exchange constants  is comparable to a relatively small hole energy anisotropic splitting. 
\end{abstract}

\maketitle
\section{Introduction}

Colloidal semiconductor nanocrystals (NCs)  are now widely studied both experimentally and theoretically due to their wide range of existing and potential practical applications \cite{Efros2021,Arquer2021,Bayer2019}. Contemporary methods of synthesis make it possible to grow NCs with desired shapes on demand. For example, cube-shaped CdSe  \cite{Lv2022}, PbSe \cite{Murray2001,NIU201238} and lead-halide perovskite \cite{Kovalenko2015} NCs and spheroidal GaAs NCs \cite{Talapin2024, Talapin2026} with a zinc-blende crystal structure  have been recently synthesized.

The studies of the band-edge exciton fine structure in semiconductor nanocrystals and quantum dots have nearly the same long history as the studies of quantum dots themselves, dating back to the  early 1990s. Experimentally, the  exciton fine structure splitting was first observed in spheroidal CdSe nanocrystals, both chemically synthesized \cite{Nirmal1995} and grown in the glass matrix \cite{Chamarro1996,Woggon1996}.

 Within the ${\bm k}\cdot {\bm p}$ method,  the effect of uniaxial anisotropy on the exciton fine structure in spheroidal NCs is usually considered  as a perturbation.   In particular, it was first shown  in Ref. \cite{Efros1996} within the first order of  perturbation theory that  the uniaxial anisotropic splitting of the light and heavy holes caused by the wurtzite  crystal field \cite{Efros1992} and by the NC spheroidal shape with small uniaxial  anisotropy \cite{Efros1993}, together with the isotropic electron-hole short-range exchange interaction, results in five band-edge exciton levels. The account of the long-range electron-hole exchange interaction contributes to the total exchange constant \cite{Goupalov1998,Goupalov2000,Sercel2018} having the same size dependence as the short-range one   and does not change the symmetry of the exciton fine structure.  Since the first experimental and theoretical research, many papers have been devoted to the study of the exciton fine structure in various semiconductor nanostructures \cite{Golovatenko2022,Bui2020,Franceschetti1999,Leung1998,Talapin2026}. Recently,  the effect of the uniaxial shape anisotropy on the long-range electron-hole interaction and the effect of the uniaxial shape anisotropy in NCs made of semiconductors with a small spin–orbit splitting of the valence band were considered theoretically \cite{Goupalov2024,Goupalov2024so}. However, all the theoretical calculations are usually done without taking into account the cubic symmetry of the semiconductor crystal lattice.

In epitaxial quantum dots, light- and heavy-hole excitons are strongly split so that the mixing of heavy and light holes via the exchange interaction with electrons is negligible. The fine structure of the ground exciton states consists of two bright (with full momentum  projection on the structure axis $\pm 1$) and two dark (with full momentum  projection $\pm 2$) states. The splitting of these doublets due to cubically symmetric exchange interaction terms in epitaxial quantum dots has been comprehensively studied both experimentally and theoretically in a number of papers. \cite{ Bayer1999, Bayer2002, Tsitsishvili2017}

In nanocrystal quantum dots, light-heavy hole splitting can be comparable to or even smaller than the electron-hole exchange interaction, resulting in a  more complicated exciton fine structure. In the latter case, cubically symmetric exchange interaction terms are usually neglected since they are much smaller  compared to isotropic exchange terms. However, as we showed recently, cubically symmetric terms caused by the valence band warping or  the cubic shape of quantum dots significantly modify the Zeeman splitting of the holes \cite{Semina2023}. An account of the cubic symmetry of the crystal lattice  and the shape of NC might be important for the description of the exciton fine structure in cubically shaped CdSe NCs as well as the new spheroidal GaAs NCs. In a recent paper \cite{Talapin2026}, the fine structure of excitons in spheroidal GaAs NCs  was studied experimentally.  

In this work, we propose  theoretical calculations of five isotropic, uniaxially anisotropic, and cubically symmetric exchange constants, as well as the light-to-heavy hole energy splitting in cuboidal and spheroidal quantum dots having an arbitrary aspect ratio beyond the first-order perturbation theory. This allows us to consistently consider the effects of  NC shape anisotropy and cubic symmetry of the crystal lattice on the fine structure of the band-edge excitons and their oscillator strength.

The paper is organized as follows: in Section \ref{spherical_states} we introduce the structures under study and the basic Hamiltonians to describe electron and hole states; in Section \ref{shape_energy} we  study the effect of  uniaxial anisotropy on the splitting of the lowest even hole state; in Section \ref{exch_general} we present the general form of the effective Hamiltonian of electron-hole exchange interaction with five constants, which are calculated in Section \ref{calc_short_range} for short-range contributions to the exchange interaction; in Section \ref{Discussion}  we discuss the possible effect of the anisotropy of the exchange interaction  constants on the exciton fine structure and oscillator strength; finally, in Section \ref{concl} we conclude our results. 

\section{Electron and hole states in nanocrystals with spherical and cubical shapes}\label{spherical_states}

We consider  zinc-blende direct-band semiconductors. Electrons near the bottom of the $\Gamma_6$ conduction band are described by the isotropic parabolic dispersion
\begin{equation}\label{He}
    \widehat{H}_e=\frac{\hbar^2}{2 m_e}k_e^2 \,
\end{equation}
We assume that the energy of the valence band spin-orbit splitting  is positive and sufficiently large compared to the hole quantization energy. So, the topmost valence subband is the  $\Gamma_8$ four-fold degenerate subband, and holes are described by the four-band Luttinger Hamiltonian:
\begin{equation}\label{lutt}
\widehat{H}_L=\frac{\hbar^2}{2 m_0}\left[\left(\gamma_1+\frac{5}{2}\gamma_2\right)k^2-2\gamma_2\sum_{\alpha}J_{\alpha}^2k_{\alpha}^2-2\gamma_3\sum_{\alpha\neq\beta}\{J_\alpha J_\beta\}\{k_{\alpha}k_{\beta}\}\right],\quad \alpha,\beta=x,y,z.
\end{equation}
Here $J_{\alpha}$ are matrices of momentum $J=3/2$, $\gamma_1, \gamma_2, \gamma_3$ are Luttinger parameters characterizing $\Gamma_8$ valence band dispersion, $x,y,z$ denote the coordinate axis along crystallographic directions, and  $\{AB\}=(AB+BA)/2 $. In the spherical approximation for the Luttinger Hamiltonian, one can replace Luttinger parameters $\gamma_2$ and $\gamma_3$ with the isotropic Luttinger constant, $\gamma=(2\gamma_2+3\gamma_3)/5$,  chosen in such a way as to obtain  zero first order correction to the hole energy with respect to the warping parameter $\alpha=(\gamma_3-\gamma_2)/\gamma$ in an otherwise spherically-symmetric system \cite{Gelmont1971, Efros1998}. In this approximation, the heavy and light hole effective masses, $m_{hh}=m_0/(\gamma_1-2\gamma)$ and $m_{lh}=m_0/(\gamma_1+2\gamma)$, are isotropic. Note that Eq. \eqref{He} and the spherical approximation of the Luttinger Hamiltonian  Eq. \eqref{lutt} are often used for the electron and hole dispersions in wurtzite semiconductors in the quasi-cubic approximation where the additional crystal field is considered as a perturbation.  

\begin{figure}
\includegraphics[width=0.95\columnwidth]{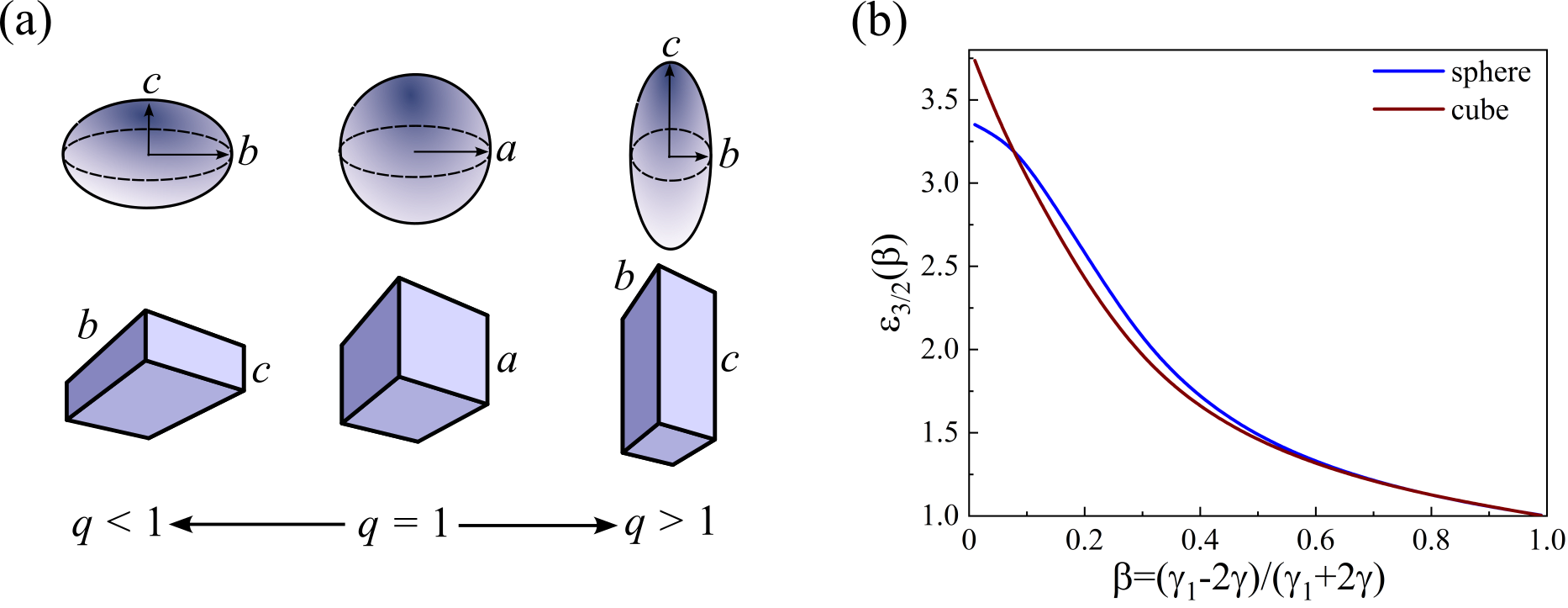}
   \caption{(a) The Sketch of the structures under study: (upper row) spherical NC with  radius $a$  and  spheroidal NCs with half-axes $b$ and $c$; (lower row) cubical NC with  edge $a$ and cuboidal NCs with edges  $b$ and $c$.  Left and write columns show the oblate NCs with $q=c/b <1$  and prolate $q=c/b >1$, respectively; (b) Dependence of the hole ground state  energy $\varepsilon_{3/2}(\beta) = E_{1S_{3/2}}/E_h$ in  spherical and  cubical NCs on the light to heavy hole mass ratio $\beta=(\gamma_1-2\gamma)/(\gamma_1+2\gamma)$ calculated in spherical approximation for Luttinger Hamiltonian.} 
 \label{NC_epsilon}
\end{figure}

We consider electrons and holes confined in the nanocrystals with spheroidal or cuboidal symmetry shown in the upper and lower rows of  Fig. \ref{NC_epsilon}(a), respectively. 
In the fully spherically or cubically symmetric case  (middle column), we denote the radius of the sphere or the edge of the cube by $a$,  while in the uniaxial structures, the half-axis of the spheroid or edges of the cuboid are denoted as $b$ (for two equal directions) and $c$ (for the anisotropy direction).    We compare NCs with the same volume so that $b^2c = a^3$ and the edge  of the cube $a \equiv a_c$ is larger than the radius of the sphere $a \equiv a_s$  by a factor of $ a_c/a_s=(4\pi/3)^{1/3}$.   The uniaxial anisotropy of the NC shape is characterized by the aspect ratio $q=c/b$.

We assume the energy potential profile of the NCs to be a box-like infinite potential so that the electron and hole wave functions  vanish at the  NC surface.
 We consider NCs with sizes $a,b,c$ smaller than the bulk exciton Bohr radius, so that the strong quantization regime for electrons and holes is realized, and the exciton wave function can be factorized into  the product of the  electron and hole wave functions. 
In this paper, we study the fine structure of the band-edge exciton comprising the ground state electron, $1S_e$  and the lowest even hole state $1S_{3/2}$, which is in most cases the hole ground state. In the exciton description, it is natural to use  electron representation for electrons and hole representation for holes  

The ground state of the electron in all considered NCs is two-fold degenerate and is characterized by the total angular momentum $1/2$ and its projections on the $z$-axis:  $\Psi_{m}^e(\bm r_e)$, $m=\pm1/2$.  For $q=1$ the energies of the $1S_e$ state in spherical, $E_e \equiv E_e^s$, and cubical, $E_e \equiv E_e^c$, NCs are $$E_e^s=\frac{\hbar^2\pi^2}{2m_e a_s^2} \equiv \frac{\hbar^2\pi^2}{2m_e a^2} , \quad E_e^c=\frac{3\hbar^2\pi^2}{2m_e a_c^2} \equiv \frac{3\hbar^2\pi^2}{2m_e a^2}.$$ 

The $1S_{3/2}$ state of quantum-confined holes  in spherically-symmetric system  is four-fold degenerate with respect to the projection of the total angular momentum $M$: $M=\pm3/2$, $\pm1/2$, and described by the
wave functions \cite{Gelmont1971,Semina2016}  
\begin{eqnarray}
\psi_{M}^h \left( {\mathbf{r}_h} \right) = 2 \left( -1\right)^{M-3/2}\sum_{l=0,2}  
i^l R_{l}\left( r_h\right)
\sum_{m+\mu = M} \left(
\begin{array}{ccc}
l & 3/2 & 3/2 \\
m &\mu &  -M
\end{array}
\right)
Y_{lm}\left( \theta,\phi\right) u_\mu
\label{def}
\end{eqnarray}
The spherical harmonics $Y_{lm}\left( \theta,\phi\right)$ are defined in \cite{Edmonds1974}. The factor $i^l$ was introduced in \cite{Semina2016} in order to have the same sign of the radial wave function $R_2$  and the same system of differential equations as derived in \cite{Gelmont1971} by using the other definition of spherical harmonics from \cite{Landau3}.
For NCs with cubic symmetry of the crystal lattice or NC shape,  the hole ground state remains four-fold degenerate while the wave functions $\Psi_{M }^h(\bm r_h)$ have a more complicated structure \cite{Semina2021,Semina2023}. Note that even a spherical NC becomes cubically-symmetric if we take into account the  cubic symmetry of the crystal lattice (valence band warping with $\gamma_3\neq \gamma_2$). For cuboidal NC, we assume that its edges  are directed along the crystallographic axes $x,y,z$.

The energy of  $1S_{3/2}$ state, $E_{1S_{3/2}}$, in the spherical approximation for the Luttinger Hamiltonian depends on the light-to-heavy hole mass ratio $\beta=m_{lh}/m_{hh}=(\gamma_1-2\gamma)/(\gamma_1+2\gamma)$. 
 The  energy $E_{1S_{3/2}}(\beta)=\varepsilon_{3/2}(\beta)E_h$ as a function of $\beta$ for spherical (with $E_h \equiv E_h^s$) and cubical NCs (with $E_h \equiv E_h^c$) is shown in  Fig. \ref{NC_epsilon}(b), where
  $$E_h^s=\frac{\hbar^2\pi^2}{2m_{hh} a_s^2} \equiv\frac{\hbar^2\pi^2}{2m_{hh} a^2}, \quad E_h^c=\frac{3\hbar^2\pi^2}{2m_{hh} a_c^2} \equiv\frac{3\hbar^2\pi^2}{2m_{hh} a^2}$$ 
   are defined with the heavy hole mass  $m_{hh}=m_0/(\gamma_1-2\gamma)$. In the limit of the simple valence band, $\beta\rightarrow 1 $, the value $\varepsilon_{3/2}(\beta)\rightarrow 1 $ for both types of NCs and increases with a decrease of $\beta$. In spherical NCs,  $\varepsilon_{3/2}(\beta\rightarrow 0) \rightarrow \phi_2^2/\pi^2\approx 3.36 $, where $\phi_2 \approx 5.76$ is the first zero of the spherical Bessel function \cite{Efros1996}. For  NCs with cubical shape,  our numerical calculations yield no finite value of $\varepsilon_{3/2}(\beta\rightarrow 0) $. 
The effect of the cubic anisotropy coming from $\gamma_2\neq \gamma_3$ or the cubical shape of the NC, as well as the uniaxial anisotropy  on the electron and  hole energy states in spheroidal and cuboidal NCs  will be discussed in the next Section \ref{shape_energy}.

\section{Electron and hole states in nanocrystals with spheroidal and cuboidal shapes}
\label{shape_energy}

 As examples of uniaxially anisotropic structures, we consider spheroidal (rotational ellipsoid) and cuboidal (rectangular parallelepiped with a square cross section) NCs shown in Fig. \ref{NC_epsilon}(a) in the left and right columns.

To consider the uniaxial anisotropy,  we introduce the NC dimension along the anisotropy axis ($z'$-axis) as $c$ and along $x',y'$ perpendicular to $z'$ as $b$. In  the general case, the axes $x',y',z'$ may not coincide with the crystallographic axes $x,y,z$. To treat the anisotropic structure, it is useful to perform the transformation of coordinates $z' \rightarrow z'(c/a)$ and $x' \rightarrow x'(b/a)$, $y' \rightarrow y'(b/a)$, which restores the spherical or cuboidal symmetry of the confinement potential $V(\bm r)$. Upon such a transformation,  the additional terms $\hat{H}_{\rm an}^e$ and $\hat{H}_{\rm an}^h$ in the electron and hole kinetic  energy Hamiltonians arise. These terms shift but  do not split two-fold degenerate electron ground s-like states,  and both shift and split the four-fold degenerate hole ground state  into two doublets with projection of the total angular momentum on the structure axis $M=\pm 3/2$ and $M=\pm 1/2$. We refer to states with $M=\pm 3/2$ as heavy holes and to states with $M=\pm 1/2$ as light holes. At small anisotropy, these states are a mixture of both; however,  the states with $M=\pm 3/2$ consist mostly of heavy holes and states with $M=\pm 1/2$  mostly of light holes, with growing anisotropy. We consider two directions of the anisotropy axis, $z' \parallel [001]$ and $z' \parallel [111]$, for spheroidal NCs and one direction,  $z' \parallel z \parallel [001]$, for cuboidal NCs. 

From  the volume conservation condition, $b^2c=a^3$, and the anisotropy parameter $q=c/b$ we obtain
\begin{equation}\label{vol}
 b=aq^{-\frac{1}{3}}\, , \, \,   c=aq^{\frac{2}{3}} \, .
\end{equation}
The definition in Eq. \eqref{vol} does not require the smallness of the anisotropy, so that  $q$ may have arbitrary values ranging from $0$ for two-dimensional (2D)  structures to $\infty$ for one-dimensional (1D) structures.

For arbitrary value of $q$ the  electron kinetic energy Hamiltonian after coordinate transformation  has the simple form:
\begin{equation}\label{electron}
\hat{H}_e=-\frac{\hbar^2}{2m_e}\left[q^{2/3}\left(\frac{\partial^2}{\partial x^2}+\frac{\partial^2}{\partial y^2}\right)+q^{-4/3}\frac{\partial^2}{\partial z^2}\right].    
\end{equation}
Electron states in cuboidal NCs can be found analytically without a coordinate transformation. It allows us to verify the results of the numerical calculations using Hamiltonian \eqref{electron} with boundary conditions of the cubical NCs with the edge $a$.  For spheroidal NCs, we  calculate electron wave functions  numerically  using the Hamiltonian \eqref{electron} with boundary conditions of the spherical NCs with  radius $a$. The  numerical method is the same  one as used for holes in Ref. \cite{Semina2023} and below.  Note that the uniaxial anisotropy can shift the energy of the electron ground states nonlinearly with respect to the anisotropy parameter $\mu = q-1$ \cite{Landau3}, but it does not result in its splitting and thus does not contribute to the exciton fine structure splitting. However, we use the electron anisotropic wave functions below to calculate the constants of the electron-hole exchange interaction for arbitrary values of $q$.    

Upon the coordinate transformation, the hole  Hamiltonian \eqref{lutt} transforms to $\hat H_L+ \hat H_{\rm an}^h$. The explicit form of $\hat H_{\rm an}^h$ depends on the direction of the anisotropic axis $z'$ with  respect to the crystallographic axes. If the  anisotropy axis $z'$ coincides with the crystal $z$-axis  ${z'|| z || [001]}$ and additionally $x'||x||[100]$, $y'||y||[010]$ the resulting symmetry of the NC would be the highest: point symmetry groups $D_{4h}$ or $D_{2d}$ depending on the existence of the inversion center in bulk crystal.

The matrix form of the $\hat{H}_{\rm {an}}^h=\hat{H}_{001}^h$ for $z'|| z || [001]$ for an arbitrary $q$ value is presented  in Eq. \eqref{lutt_matrix_an} in Appendix \ref{AA}. 
In the limit of small anisotropy parameter $$\mu=q-1 = c/b-1 \, $$
in the  linear-in-$\mu$ regime, we arrive at the  following correction to the hole Hamiltonian:
\begin{equation}\label{Han}
\hat{H}_{001}^{\mu}=\frac{2\mu}{3}\hat{H}_L-\frac{\hbar^2}{2 m_0}2\mu\left[\left(\gamma_1+\frac{5}{2}\gamma_2\right)k_z^2-2\gamma_2k_z^2J_z^2-2\gamma_3\{(k_xJ_x+k_yJ_y)k_zJ_z\}\right].
\end{equation}
We note that for different definitions of  $\mu$ (see, for example,  
 Refs. \cite{Migdal1977,Goupalov2024,Efros1993,Semina2016}),  the resulting perturbation Hamiltonian obtained in linear on $\mu$ regime is the same as Hamiltonian     \eqref{Han}.

We consider now  the case of the anisotropic axis  $z'$  along [111] direction ($x' \parallel$ [11-2] and $y'\parallel $[-110]) as reported for newly synthesized  GaAs NCs \cite{Talapin2024}. The point symmetry group would be $D_{3d}$ or $C_{3v}$ depending on the existence of inversion symmetry in the bulk crystal. To derive the anisotropic correction to the hole kinetic energy  $\hat H_{111}^h$ we   use the relation between the axis $x,y,z$ and $x',y',z'$:
\begin{equation}
x= \frac{x'}{\sqrt{6}} - \frac{y'}{\sqrt{2}}+\frac{z'}{\sqrt{3}} \, , \quad 
y= \frac{x'}{\sqrt{6}} + \frac{y'}{\sqrt{2}}+\frac{z'}{\sqrt{3}} \, ,\quad 
z= -\frac{\sqrt{2}}{\sqrt{3}}x' +\frac{z'}{\sqrt{3}} \,  
\end{equation}
and perform the following coordinate transformation in the Luttinger Hamiltonian:
\begin{eqnarray}
x \rightarrow   \frac{(2 b+c)}{3 a}x-\frac{(b-c)}{3 a}y-\frac{(b-c)}{3 a}z\, , \\ 
y \rightarrow -\frac{(b-c)}{3 a}x+\frac{(2 b+c)}{3 a}y-\frac{(b-c)}{3 a}z\, ,\\
z\rightarrow -\frac{(b-c)}{3 a}x-\frac{(b-c)}{3 a}y+\frac{(2 b+c)}{3 a}z\, . 
\end{eqnarray}
For an arbitrary $q$, the correction to the hole Hamiltonian $\hat{H}_{111}^h$  is shown in Appendix \ref{AA}, Eq. \eqref{lutt_matrix_an111}. In the limit of small anisotropy, we obtain 
\begin{multline}\label{Han111}
\hat{H}_{111}^{\mu}=-\frac{\hbar^2}{2 m_0}\frac{2}{3}\mu[\left(\gamma_1+\frac{5}{2}\gamma_2\right)\sum_{\alpha\neq\beta}k_{\alpha}k_{\beta}-\gamma_2\sum_{\alpha\neq\beta\neq\gamma}k_{\alpha}k_{\beta}\left(\frac{15}{4}-J_{\gamma}^2\right)-\\-\gamma_3\sum_{\alpha\neq\beta\neq\gamma}k_{\alpha}^2\{J_{\alpha}\left(J_{\beta}+J_{\gamma}\right)\}-\gamma_3\sum_{\alpha\neq\beta\neq\gamma}k_{\alpha}k_{\beta}\{J_{\gamma}\left(J_{\alpha}+J_{\beta}\right)\}].
\end{multline}

 If the spheroidal NC (spherical after the coordinate transformation) is considered in spherical approximation for the Luttinger Hamiltonian~\eqref{lutt} with  $\gamma_2$ and $\gamma_3$ replaced by the isotropic Luttinger parameter $\gamma=(2\gamma_2+3\gamma_3)/5$, the effect of the uniaxial anisotropy on the  hole ground state energy    does not depend on the direction of the anisotropic axis $z'$ and the corrections coming from $\hat{H}_{001}^h$ and from  $\hat{H}_{111}^h$ are the same. 

\begin{figure}
\includegraphics[width=0.95\columnwidth]{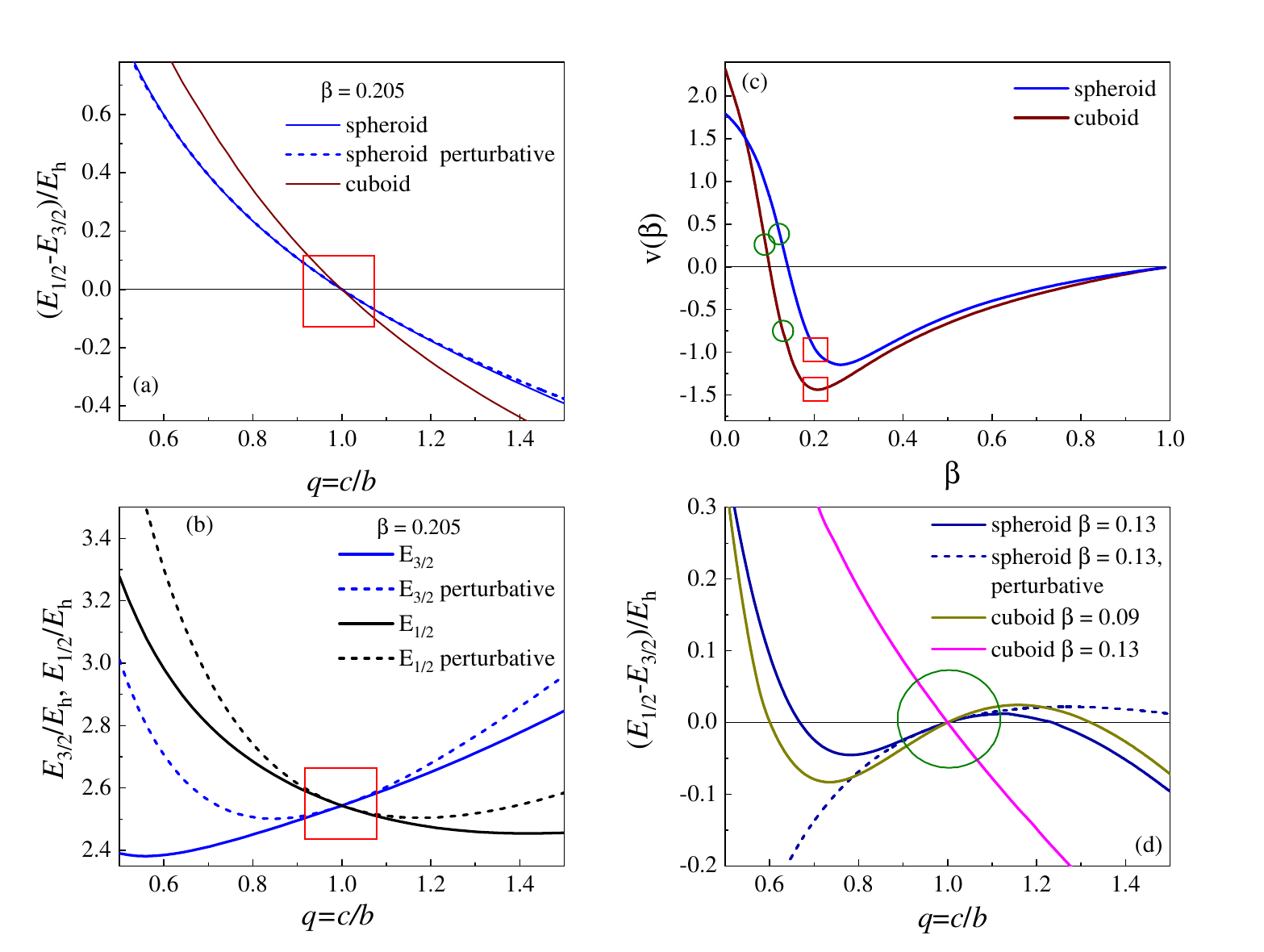}
   \caption{(a) The  heavy and light hole splitting $\Delta = E_{1/2} - E_{3/2}$ in units of $E_h$ as a function of $q=c/b$. Solid  lines correspond to results of numeric calculation, dashed line corresponds to a perturbative treatment of Hamiltonian \eqref{lutt_matrix_an} with allowance for an arbitrary value of $q$. Calculations were made for zb-CdSe valence band parameters in the spherical approximation for Luttinger Hamiltonian \cite{Fu1998n2} with $\beta=0.205$; (b) Energies of heavy, $E_{3/2}$, and light, $E_{1/2}$,  holes in units of $E_h$ calculated for spheroid NCs with CdSe parameters, $\beta=0.205$. Solid lines correspond to full numerical calculation,  dashed lines correspond to perturbative treatment of Hamiltonian \eqref{lutt_matrix_an} with allowance of arbitrary value of $q$; (c)  the slope of heavy and light hole splitting at small anisotropy $v(\beta)$ as a function of the heavy to light hole mass ratio $\beta=(\gamma_1-2\gamma)/(\gamma_1+2\gamma)$;    (d) heavy and light hole splitting in units of $E_h$  as a function of ratio $q$ for $\beta$, corresponding to positive $v(\beta)$:  $\beta=0.13$ (GaAs parameters)  for  spheroidal NCs and  $\beta=0.09$ and $\beta=0.13$ for cuboidal NCs. The dashed line shows the result of perturbative treatment of the Hamiltonian \eqref{lutt_matrix_an} with allowance of an arbitrary value of $q$. Red squares and green circles in panels (a), (b) and (d) represent the range of $q$ values where the linear regime is realized. Squares and circles in panel (c) denote the slopes of the splittings in the linear regime shown in panels (a) and (d), respectively. }   
 \label{anis_en1}
\end{figure}

For further description of the exciton fine structure, we will be mostly interested in the anisotropy energy splitting between the light ($|M|=1/2$) and heavy ($|M|=3/2$) hole states, 
$\Delta = E_{1/2} - E_{3/2}$,
in the NCs with uniaxial anisotropy. This splitting  can be described by the effective Hamiltonian:
\begin{equation}\label{H_delta}
\hat{H}_{an}= \frac{\Delta}{2}\left(\frac{5}{4}-J_{z'}^2\right).
\end{equation}
In the case of small anisotropy, the energy splitting $\Delta$ can be found within the framework of perturbation theory by taking the diagonal matrix elements of the $\hat{H}_{001}^{\mu }$ \eqref{Han} or  $\hat{H}_{111}^{\mu}$ \eqref{Han111} perturbations at the $\Psi_{M=\pm 3/2}$ and $\Psi_{M=\pm 1/2}$ ground state wave functions of the $\hat H_L$ Hamiltonian for spherical or cubical NCs. 
Such a  splitting $\Delta $ in the form  
$$\Delta=\mu u(\beta)E_{S_{3/2}}(\beta)$$  was  calculated first  by using perturbation theory with analytically found wave functions for the spherical NCs with a box-like potential in  Ref. \cite{Efros1993}. The dimensionless function $u(\beta)$ depends only on the light-to-heavy hole mass ratio $\beta=(\gamma_1-2\gamma)/(\gamma_1+2\gamma)$ via the integrals of the radial wave function combinations (the explicit expressions  can be found in Ref. \cite{Semina2016}).

In this work, we go beyond  perturbation theory and calculate numerically the hole energy splitting $\Delta$ for any value of $q$.  We made our illustrative calculations for one set of zb-CdSe band parameters \cite{Fu1998n2}: $\gamma_1=2.52$, $\gamma_2=0.65$, $\gamma_3=0.95$ (in spherical approximation   $\gamma=0.83$) and $\beta=0.205$ and for one set of GaAs parameters \cite{Baldereschi1973}: $\gamma_1=7.65$, $\gamma_2=2.41$, $\gamma_3=3.28$ ($\gamma=2.93$) and $\beta=0.13$.

In Fig. \ref{anis_en1} (a), we  illustrate the applicability of the perturbative treatment of the Hamiltonian \eqref{lutt_matrix_an}. In the figure, we show  the results of  the calculation of the hole lowest symmetric state  energy splitting  in units of  $E_h$ for volume conservation condition for zb-CdSe valence band parameters  as a function of $c/b$. Calculations were made for spheroidal and cuboidal NCs with a box-like potential in the spherical approximation for the Luttinger Hamiltonian.  
Solid  lines correspond to the  results of numerical calculations; dashed lines correspond to the perturbative treatment of Hamiltonian \eqref{lutt_matrix_an} by analogy with Ref. \cite{Efros1993}, with allowance for an arbitrary value of $q=c/b$ for spheroidal NCs (for cuboidal NCs the simple perturbative treatment is not possible). That means that while treating Eq. \eqref{lutt_matrix_an} perturbatively, we do not expand it up to the linear on $\mu=q-1$ terms. This allows us to obtain a better agreement between the numerical and perturbative calculations for larger anisotropy. By the red square, we show the range of the $q$ values where the splitting $\Delta$ is a linear function of  the small anisotropy parameter $\mu=c/b-1$.   One can see that for the hole ground state energy splitting, the perturbation theory without linearizing on anisotropy parameters gives very good agreement with numerical calculations even outside the linear regime.  Although panel (b) of  Fig. \ref{anis_en1} shows that the hole energy levels $E_{3/2}$ and $E_{1/2}$ themselves can be described by extended perturbation theory only in the range of the anisotropy parameter $q$ where the linear regime is realized (denoted by a red square).

Fig. \ref{anis_en1} (c)  shows the slope of heavy and light hole splitting characterized by the function $v(\beta)=u(\beta) E_{S_{3/2}}(\beta)/E_h$ at small anisotropy, where the linear regime is realized: $$\Delta=\mu v(\beta)E_h  \equiv \mu u(\beta) E_{S_{3/2}}(\beta)/E_h$$  calculated in the spherical approximation for the Luttinger Hamiltonian.  Remarkably, for spheroidal   NCs  at $\beta\approx 0.14$ and for cuboidal NCs at $\beta\approx 0.1$, the quantity $v(\beta)$ changes sign.  For positive values of $v(\beta)$ in some range of $c/b$, the ordering of the hole levels is  counterintuitive:   in oblate NCs  the hole ground state  would be the light hole with projection $M=\pm 1/2$ in contrast to the situation in $2D$ quantum wells, and in prolate NCs the ground state would be the heavy hole with projection $M=\pm 3/2$ in contrast to the situation in $1D$ quantum wires. Note that the quantity $v(\beta)$ for spheroidal NCs with parabolic potential, calculated in Ref. \cite{Semina2016}, has the same sign for all values of $\beta$.

 In Fig. \ref{anis_en1} (d), we show the light and heavy hole splitting  in  units of $E_h$ calculated for spheroidal and cuboidal NCs  for values of $\beta$ corresponding to the range of  positive $v(\beta)$: $\beta=0.09$ for cuboidal NCs, and $\beta=0.13$ for spheroidal NCs. One can see that because $v(\beta)>0$, the  order of hole states at small anisotropy is inverted: in slightly oblate NCs  the hole ground state is the  hole state with $M=\pm 1/2$, and in slightly prolate NCs the ground state is the  hole state with $M=\pm 3/2$. For such values of $\beta$, the splitting becomes a non-monotonic function of the anisotropy parameter  $q$, even changing its sign. At larger anisotropy, the order of hole states becomes normal: in the limit of planar NC the ground state is always the  hole  state with $M=\pm 3/2$, and in the limit of the quantum wire the ground state is always the  holes with $M=\pm 1/2$. Importantly, for several semiconductors, for example GaAs or CdTe with $\beta \approx 0.13$,  the anisotropic splitting  $\Delta$ in spheroidal NCs can be zero for substantial uniaxial anisotropy of the NC shape. It is worth noting that the perturbation treatment in this situation works well only in the range of  the linear-in-anisotropy regime  shown by the green circle, even for the level splitting; see the dashed line. In contrast, for cuboidal NCs with $\beta=0.13$ we obtain  $v(\beta)<0$ and, as one can see from  Figure \ref{anis_en1}(d), the dependence of hole state splitting on the anisotropy parameter remains monotonic.

\begin{figure}[h]
\includegraphics[width=0.95\columnwidth]{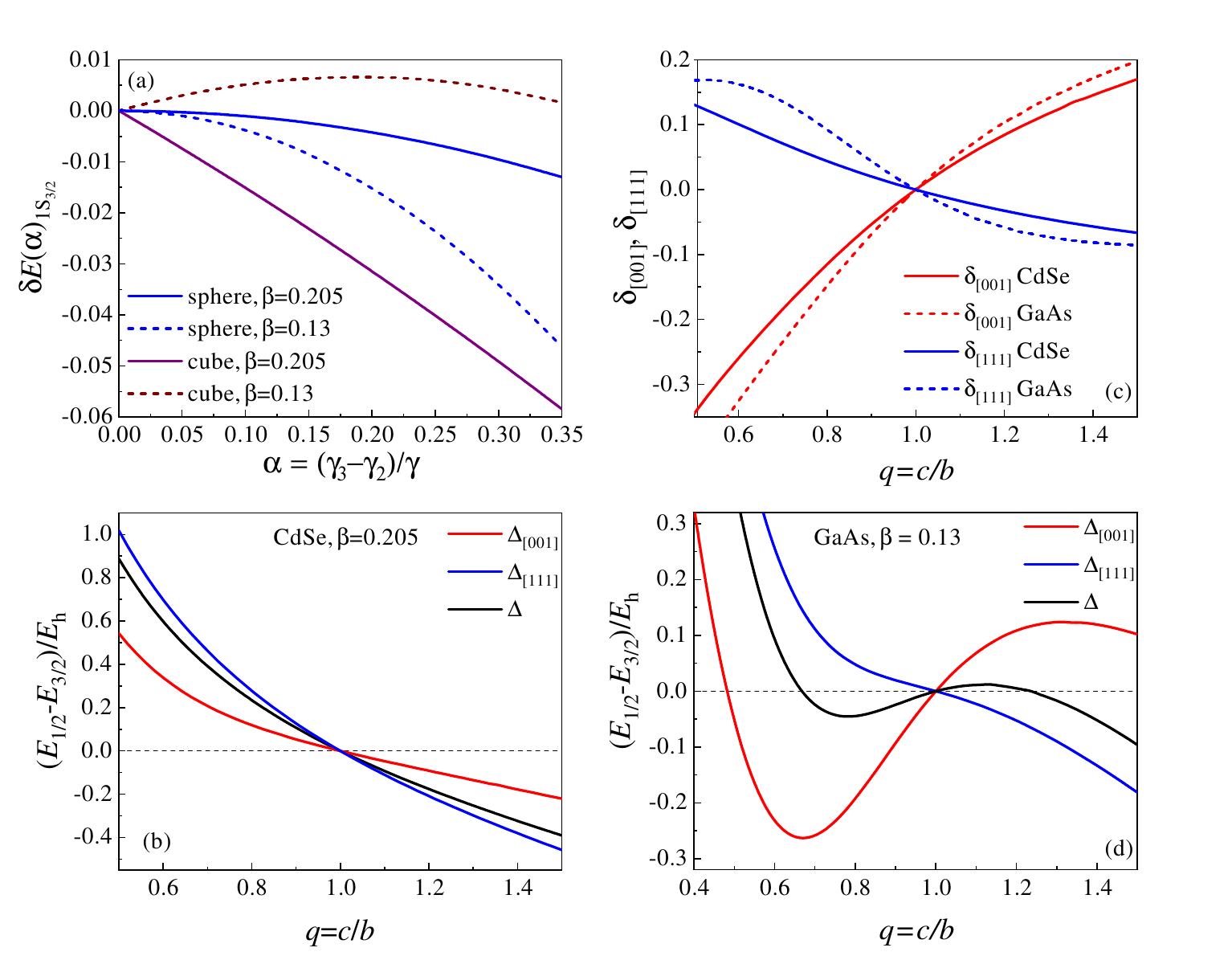}
   \caption{  (a) The relative accuracy of the spherical approximation for Luttinger Hamiltonian $\delta E_{1S_{3/2}}(\alpha)=(E_{1S_{3/2}}(\beta,\alpha)-E_{1S_{3/2}}(\beta,0))/E_{1S_{3/2}}(\beta,0))$,  calculated for spherical and  cubical NCs calculated for  $\beta=0.205$  (solid lines) and $\beta=0.13$ (dashed lines);  (b,d) The hole $1S_{3/2}$ state splitting $\Delta = E_{1/2} - E_{3/2}$ in units of $E_h$ calculated as function of anisotropy parameter $q$ for spheroidal (b) CdSe NCs   and  (d) GaAs NCs  in spherical approximation for the Luttunger Hamiltonian  and with account of the cubic symmetry of the crystal lattice:  $\Delta_{[001]}$ for the anisotropy axis $z'$ along $[001]$ and $\Delta_{[111]}$  for $z'$ along $[111]$   directions;  (c) The relative effect of the valence band warping   on the hole ground state splitting in spheroidal  NCs $\delta_{[001]} =(\Delta_{[001]} - \Delta)/E_h$ and $\delta_{[111]} =(\Delta_{[111]} - \Delta)/E_h$ as a function of the anisotropy parameter $q=c/b$. } 
 \label{spher_app}
\end{figure}

 If the valence band warping $\gamma_2\neq\gamma_3$ is included, the $1S_{3/2}$  hole state in spherical or cubical NCs remains four-fold degenerate.  The  energy of hole $1S_{3/2}$ state, $E_{1S_{3/2}}(\beta,\alpha)$,  becomes a function of two parameters, $\beta$ and the warping parameter $\alpha$. In Fig.  \ref{spher_app} (a) the relative corrections $\delta E_{1S_{3/2}}(\alpha)=(E_{1S_{3/2}}(\beta,\alpha)-E_{1S_{3/2}}(\beta,0))/E_{1S_{3/2}}(\beta,0))$ calculated for spherical and cube NCs for $\beta=0.205$ (CdSe) and $\beta=0.13$ (GaAs) are shown. One can see that in spherical NCs the linear-in-$\alpha $ corrections, $\Delta_{\text{warp}}(\alpha)$, are indeed absent at small values of $\alpha $. In contrast,  for cube NCs, the linear corrections are present, although they are rather small. For realistic values of the warping parameter, $\alpha \approx 0.36$ ($0.3$) for our CdSe (GaAs) set of parameters, the relative corrections are no larger than 6 percent.

We denote the hole $1S_{3/2}$ state splitting calculated in the spherical approximation for the Luttinger Hamiltonian as $\Delta$ and the hole ground state splitting calculated with account for the cubic symmetry of the crystal lattice (valence band warping) as $\Delta_{[001]}$  and  $\Delta_{[111]}$ for the   anisotropy axis $z'$ directed along $[001]$ and $[111]$, respectively.     In Fig. \ref{spher_app} (b) and (d)  we show the $\Delta_{[001]}$, $\Delta_{[111]}$ and $\Delta$    calculated for spheroidal NCs  with CdSe and GaAs parameters, respectively.  One can see the significant effect of the account of the valence band warping on hole $1S_{3/2}$ state splitting, especially for GaAs: if the anisotropic axis $z'$ is directed along  $[111]$ in GaAs NCs the light-to-heavy hole splitting $\Delta_{[111]}$ becomes the monotonic function of the anisotropy $q$ in contrast to $\Delta_{[001]}$ for $z' \parallel [001]$ or $\Delta$  in the case if the cubic symmetry is neglected.  That indicates that one should be very careful when calculating the hole ground state splitting for semiconductors  with $\beta$ close to $0.14$: the resulting value and even the sign of the splitting would be quite sensitive to the particular parametrization of the valence band parameters and to the direction of the anisotropy axis.  While the dependence of $\Delta$ on $q$ looks very different for CdSe and GaAs, see (Fig. \ref{spher_app} (b) and (d)), the corrections due to the valence band warping  $\delta_{[001]} =(\Delta_{[001]} - \Delta)/E_h$ and $\delta_{[111]} =(\Delta_{[111]} - \Delta)/E_h$ to the splitting originating from the account of the cubic symmetry of the crystal lattice are very similar (see  Fig. \ref{spher_app} (c)).

Thus, uniaxial anisotropy splits the hole states with projections $M=\pm 3/2$ and $M=\pm 1/2$ on the anisotropy axis. The value of the splitting can be nonmonotic function of the anisotropy parameter $q$ and even changes its sign (the order of the levels) depending on the direction of the anisotropy axis and light to heavy hole mass ratio $\beta$. In special cases, the anisotropic splitting might be close to zero even for $q\ne 1$, similar to the known effect in wz-CdSe prolate NCs, where the shape anisotropy compensates the effect of the crystal field \cite{Efros1996}. In all  cases of small anisotropy splitting of the hole state, it is important to study the effect of  uniaxial anisotropy on the electron-hole exchange interaction.

\section{electron-hole exchange interaction: symmetry consideration}\label{exch_general}

To study the effects of the NC shape anisotropy and/or cubic symmetry of the crystal lattice on the exciton fine structure, we first perform  the general symmetry consideration of the electron-hole exchange interaction Hamiltonian for  $D_{4h}$ or $D_{2d}$   point symmetry groups. We note that the symmetry properties of the exchange Hamiltonian are the same for    long-range and short-range contributions.

Taking into account the transformation properties of Pauli matrices $\sigma_{\alpha}$ and matrices of angular momentum $J=3/2$, $J_{\alpha}$ in $D_{4h}$ or $D_{2d}$ point symmetry groups \cite{Koster1969} we write the effective exchange interaction Hamiltonian as follows: 
\begin{eqnarray}\label{cuboid}
    \hat{H}_{\rm exch}^{\rm c+un}  = &&-\eta_\parallel \left( \sigma_z  J_z\right) -    \eta_\perp \left( \sigma_x  J_x+ \sigma_y  J_y\right)  \\
   && +\eta_\parallel^c(\sigma_zJ_z^3)+\eta_\perp^c(\sigma_x J_x^3+\sigma_yJ_y^3)+\tilde{\eta}(\sigma_x\{J_xJ_z^2\}+\sigma_y\{J_yJ_z^2\}) \, , \nonumber
\end{eqnarray}
where $\eta_\parallel, ~\eta_\perp,~ \eta_\parallel^c,~\eta_\perp^c,~ \tilde{\eta}$ are five independent constants containing both short- and long-range contributions.  

In a spherically-symmetric system, the exciton ground state $E_{ex}$ neglecting the electron-hole exchange interaction, is eight-fold degenerate. 
The exciton spin states, taking into  account the electron-hole exchange interaction  
 \eqref{cuboid},  and the effective uniaxial anisotropic  Hamiltonian  Eq. \eqref{H_delta} are the eigenstates of the effective spin Hamiltonian: 
 \begin{equation} \label{spin}
     \hat H_{\rm spin} =  \hat H_{\rm an} + \hat H_{exch}^{\rm c+un} \, .
 \end{equation}
 with the energy origin at $E_{ex}$. In general, the six resulting exciton energy levels are described by:
\begin{eqnarray}\label{fine}
 &&E_{FX}= - \frac{3}{2}\eta_\parallel - \frac{\Delta}{2} + \frac{27}{8}\eta_\parallel^c + \frac{3}{2} \eta_\perp^c~ , ~   E_{FY}= - \frac{3}{2}\eta_\parallel  - \frac{\Delta}{2} + \frac{27}{8}\eta_\parallel^c - \frac{3}{2} \eta_\perp^c~ \nonumber \\ \label{cuboid_en}
 &&E_0^{U,L}= \pm 2 \eta_\perp +\frac{\eta_\parallel }{2} + \frac{\Delta}{2} - \frac{1}{8}\eta_\parallel^c \pm  5\eta_\perp^c \mp \frac{\tilde \eta}{2} \, , \\
&& E_{+1}^{U,L}= E_{-1}^{U,L}=\frac{\eta_\parallel}{2} - \frac{13}{8}\eta_\parallel^c \pm \sqrt{f^2+d} \, . \nonumber
\end{eqnarray}
with the respective exciton wave functions:
\begin{equation}\label{cuboid_func}
\begin{gathered}
\Psi_{FX} = \frac{1}{\sqrt{2}} \left(  \Psi_{\uparrow,3/2}+ \Psi_{\downarrow,-3/2}  \right)~, 
\Psi_{FY} = \frac{1}{\sqrt{2}} \left(  \Psi_{\uparrow,3/2}- \Psi_{\downarrow,-3/2}  \right)~,\\
\Psi_0^{U,L} =  \frac{1}{\sqrt{2}}\left( \Psi_{\uparrow,-1/2} \mp \Psi_{\downarrow, 1/2} \right)\\
\Psi_1^{U} = C^{+} \Psi_{\uparrow,1/2} - C^{-} \Psi_{\downarrow,3/2}~, 
\Psi_1^{L} = C^{-} \Psi_{\uparrow,1/2} + C^{+} \Psi_{\downarrow,3/2}~, \\ 
\Psi_{-1}^{U} = C^{-} \Psi_{\uparrow,-3/2} - C^{+} \Psi_{\downarrow,-1/2}~, 
\Psi_{-1}^{L} = C^{+} \Psi_{\uparrow,-3/2} + C^{-}\Psi_{\downarrow,-1/2}~,
\end{gathered}
\end{equation} 
Here
\begin{equation} \label{df}
f=-\eta_\parallel+ \frac{7}{4}\eta_\parallel^c +\frac{\Delta}{2} = -\bar \eta_\parallel + \Delta/2,  \,  \quad d=3\left(\eta_\perp-\frac{7}{4}\eta_\perp^c-\frac{5}{4}\tilde{\eta}\right)^2=3\bar \eta_\perp^2 ,  
\end{equation}
and 
\begin{equation}\label{Cpm}
C^\pm =\sqrt{ \frac{\sqrt{f^2+d}\pm f}{2\sqrt{f^2+d}}}~~.
\end{equation}
The  $\uparrow$ denotes electron spin projection  $+1/2$ on the $z$-axis and $\downarrow$ denotes electron spin  projection $-1/2$, $\pm 3/2$ or $\pm 1/2$  are projections of the  hole full momentum.  Here the  upper index $U (L)$ corresponds to exciton state originating from triplet with full momentum $\mathcal F=1$ (five-fold degenerate state with $\mathcal F=2$).   The lower indices $+1,-1,0$ denotes the exciton state full momentum projection on $z$-axis $\mathcal F_z$. The lower index $FX$ ($FY$) corresponds to the symmetric (antisymmetric) combination of the exciton state with full momentum $\mathcal F=2$ and its z-projections $\mathcal F_z=2$ and $\mathcal F_z=-2$.

In general, the exciton fine structure and  oscillator strength depend on the relation between constants $\eta_\parallel, ~\eta_\perp,~ \eta_\parallel^c,~\eta_\perp^c,~ \tilde{\eta}$ and $\Delta$. While three upper (by energy) exciton states $0^U$ and $\pm 1^U$ originate from the state with the total angular momentum ${\cal F}=1$ and are always bright, the other five exciton states  originate from ${\cal F}=2$ and are dark unless mixed with the bright states.   Importantly, the uniaxial anisotropy not only splits the exciton levels  but also mixes the states with total angular momentum  projections $F_z=\pm 1$ originating from bright, upper, and dark, lower, excitons. As a result, the lower exciton states $\pm 1^L$ are activated and become optically-active. In spite of the many parameters affecting the exciton fine structure splittings,  the relative oscillator strength of the lower and upper excitons is controlled only by the $C^{\pm}$ constants, Eq. \eqref{Cpm},  entering the wave functions. The relative  oscillator strength $f_{\pm 1}^{U,L}$ of the bright excitons with $F_z \pm1$ with respect to the upper bright exciton with $F_z=0$ in the dipole approximation is given by 
\begin{eqnarray}
f_{\pm 1}^U/f_0^U = \frac{3}{8} (C^+ + \sqrt{3} C^-)^2 \, , \\
f_{\pm 1}^L/f_0^U = \frac{3}{8} (\sqrt{3}C^+ -  C^-)^2 \, ,
\end{eqnarray}
and 
\begin{equation}\label{ff}
f_{\pm 1}^L/f_{\pm 1}^U = (\sqrt{3} -  \tilde{f})^2/(1 + \sqrt{3} \tilde{f} )^2 \, , 
\end{equation}
where $$ \tilde{f} = \frac{C^-}{C^{+}} =\frac{\sqrt{d}}{\sqrt{f^2+d} + f}  .$$

Now we consider the special cases of the Hamiltonian Eq. \eqref{spin}, in which the overall picture is simplified. 
The simplest isotropic case with full spherical symmetry is realized for spherical NCs in the spherical approximation for the Luttinger Hamiltonian. In this case $\eta_\parallel=\eta_\perp=\eta$, $\eta_\parallel^c=\eta_\perp^c=\tilde{\eta}= \Delta=0$. The   isotropic case  was studied in detail in, for example, Refs. \cite{Efros1996,Goupalov2000}. The Hamiltonian  \eqref{spin} takes the form:
\begin{equation} \label{Hexch_sph}
\hat H_{\rm spin} =  \hat H_{\rm exch}^{\rm sph}=-\eta({\bm \sigma}_{\rm e} {\bm J} ) \,  ,
  \end{equation}
and the exciton fine structure is described by $f=-\eta, d=3\eta^2=3f^2, C^+=1/2, C^-=\sqrt{3}/2$, and $\tilde f = \sqrt{3}$ so that $f_{\pm 1}^L=0$.
 The exciton state is split into the triplet state with full momentum $\mathcal F =1$ (bright state) with the energy $E^U_{\pm 1}=E_{0}^U = E_{\rm ex} +5\eta/2$  and the five-fold degenerate state with $\mathcal F =2$ (dark state) with the  energy $E_{FX}=E_{FY}= E_{\pm 2} = E_{\pm1}^L=E_{0}^L = E_{\rm ex} -3\eta/2 $.  In all known semiconductors, the exchange energy $\eta$ is positive, so the ground exciton state is the ${\cal F}=2$ dark state.

If we consider NCs with the cubical symmetry, for example, of cubical shape grown along crystal axes or  spherical NCs made from cubic-symmetry semiconductor, there is an additional nonzero parameter of the exchange interaction $\eta_\parallel^c=\eta_\perp^c=\Delta_c$ while $\tilde \eta =0$. The Hamiltonian \eqref{spin} takes the form:
\begin{equation}\label{Hcubic}
\hat H_{\rm spin} =\hat{H}_{\rm exch}^{\rm c}=-\eta(\bm \sigma \bm J)+\Delta_c\left(\sigma_xJ_x^3+\sigma_yJ_y^3+\sigma_zJ_z^3\right) \, .
\end{equation}
The only modification of the exciton fine structure in cubically symmetric case  with respect to the spherically symmetric structure comes from  $f=-\eta+\frac{7}{4}\Delta_c$ and  $d=3\left(\eta-\frac{7}{4}\Delta_c\right)^2.$ Again, $d=3f^2$ and  $\tilde f = \sqrt{3}$ so that $f_{\pm 1}^L=0$.  The bright exciton wave functions are not modified,  but  the bright exciton energy is shifted to 
$E^U_{\pm 1}=E_{0}^U = E_{\rm ex} + \frac{5}{2}\eta - \frac{41}{8}\Delta_c$.
 At the same time,  the cubically symmetric exchange interaction $\propto \Delta_c$ mixes dark exciton states with $\mathcal F_z=2$ and $\mathcal F_z=-2$ and splits dark exciton state ${\cal F}=2$ forming the doublet with energy $E_{FX}=E_{\pm 1}^L = E_{\rm ex}- \frac{3}{2} \eta + \frac{39}{8} \Delta_c$ and the triplet with energy $E_{FY}=E_{\pm 1}^L = E_{\rm ex}- \frac{3}{2} \eta + \frac{15}{8} \Delta_c$.
The order of exciton levels depends on the sign of $\Delta_c$ and  the ratio between $\eta$ and $\Delta_c$. Our analysis shows (see  Section \ref{short} ), that typically $|\eta|\gg |\Delta_c|$ and in all our calculations we obtained $\Delta_c>0$.

In the uniaxially symmetric case realized in spheroidal NCs with neglect of the valence band warping $\eta_\perp^c=0$. The spin Hamiltonian is described by $\Delta$ (which includes all possible contributions from the shape anisotropy)  and  four independent constants of the exchange interaction with:
\begin{eqnarray}\label{uniax_exch}
    \hat{H}_{\rm exch}^{\rm un}  = &&-\eta_\parallel \left( \sigma_z  J_z\right) -    \eta_\perp \left( \sigma_x  J_x+ \sigma_y  J_y\right) +\eta_\parallel^c(\sigma_zJ_z^3)+\tilde{\eta}(\sigma_x\{J_xJ_z^2\}+\sigma_y\{J_yJ_z^2\}) \, . 
\end{eqnarray}
Similar to  the simplified spin Hamiltonian \cite{Efros1996,Goupalov2000}
\begin{equation} 
\hat H_{\rm spin} = \hat H_{\rm an}+ \hat H_{\rm exch}^{\rm sph}\,  ,
  \end{equation}
the uniaxially symmetric spin Hamiltonian $\hat{H}_{\rm exch}^{\rm un}$ of Eq. \eqref{uniax_exch}, taking into account the electron-hole exchange interaction anisotropy, splits the ground exciton state into five energy sublevels characterized by  the absolute value of the full momentum projection on the anisotropy axis even if $\Delta \approx 0$.  Moreover, one can see from Eq. \eqref{df} that even if $\Delta \approx 0$, the  activation of the $\pm 1^L$ states with $f_{\pm 1}^L \ne 0$ ($\tilde f \ne \sqrt{3}$) can be realized due to  the anisotropy of the exchange constants $\bar \eta_\parallel \ne \bar \eta_\perp$. Such anisotropy can be contributed by the anisotropy of scalar constants, $\eta_\parallel \ne \eta_\perp$,  or cubic constants, $\eta_\parallel^c \ne \tilde \eta_\perp^c$. We denote in the general case $\tilde \eta_\perp^c =  \eta_\perp^c+ 5\tilde \eta/7$ when the cubic anisotropy is taken into account together with the uniaxial anisotropy.

In  Section \ref{short}, we analyze numerically the uniaxial anisotropy and cubic-symmetry contribution of the short-range electron-hole exchange constants, while in  Section \ref{Discussion}, we analyze numerically  the impact of the uniaxial exchange constants  anisotropy on the activation of the $\pm 1^L$  exciton states.

\section{Calculation of short-range interaction constants}\label{calc_short_range}
\label{short}

The long-range exchange interaction in axially-symmetric NCs beyond the isotropic approximation was studied in Ref. \cite{Goupalov2024}. The calculations were done within the first order of perturbation theory in the linear on small anisotropy parameter regime. In contrast to the long-range exchange interaction, the short-range part remains isotropic in this limit.
In this Section, we calculate the short-range constants of effective Hamiltonians for spheroidal and cuboidal NCs with arbitrary anisotropy.

The  short-range electron-hole exchange interaction in the bulk is described by the Hamiltonian
\begin{equation} \label{hexch0}
{\cal H}_{\rm exch} = - w\hspace{0.5 mm}{\bm \sigma}^e \cdot {\bm \sigma}^h   \delta({\bm r}_e - {\bm r}_h)\:,\:  
\end{equation}
where $w=w_0 \Omega_0$, $w_0$  is an energy  exchange constant and $\Omega_0$ is the volume of the unit cell, $\sigma_{\alpha}^e$ and $\sigma_{\alpha}^h$ ($\alpha=x,y,z$) are the Pauli matrices acting on the spin  of the electron and hole, respectively. 
For the localized exciton (electron-hole pair), the short-range interaction  can be calculated by averaging the Hamiltonian \eqref{hexch0} with the exciton wave functions. The resulting effective Hamiltonian would inherit the symmetry of the structure.  Note that the bulk cubic symmetry exchange term 
\begin{equation} \label{hexch1} 
{\cal H}_{\rm exch}^c =  w_c \Omega_0 \hspace{0.5 mm}{\bm \sigma}^e \cdot {\bm j}^3   \delta({\bm r}_e - {\bm r}_h)\:,\:  
\end{equation}
is also present in bulk semiconductors with a cubic crystal lattice. However, we assume the cubic bulk exchange constant  $w_c \ll w_0$ and neglect it hereafter.

We treat the short-range exchange interaction of Eq. \eqref{hexch0} as a perturbation and consider small NCs, where the strong size quantization regime  for the  exciton is realized.   In that case, the exciton wave function is factorized into  the electron and hole wave functions as:
\begin{equation}\label{Psi_ex}
    \Psi_X(\bm r_e, \bm r_h)=\Psi_e(\bm r_e)\Psi_h(\bm r_h)
\end{equation}
To calculate the constants of the short-range exchange interaction effective Hamiltonian, we average the perturbation \eqref{hexch0}  with the wave function \eqref{Psi_ex} calculated by methods developed earlier for spherical NCs with rectangular infinite and parabolic potential, and for cubical NCs with rectangular infinite potential \cite{Semina2015,Semina2016, Semina2021, Semina2023}. Although the Hamiltonian \eqref{hexch0} is spherically-symmetric, the resulting correction will inherit the symmetry of the exciton wave function. For example,  for a cubically symmetric system with a cubically symmetric wave function \eqref{Psi_ex}, the exchange interaction effective Hamiltonian will also  be cubically symmetric. For the spheroidal and cuboidal NCs we use the numerically found wave functions with the respective symmetry.  The advantage of our numerical method is the ability to go beyond the limits of applicability of perturbation theory, which lies in the range of the anisotropy parameter $q$ approximately $q\approx 0.9\div1.1$.

\subsection{In nanocrystals with spherical and cube shapes}

\begin{figure}[h!]
    \centering
    \includegraphics[width=0.98\linewidth]{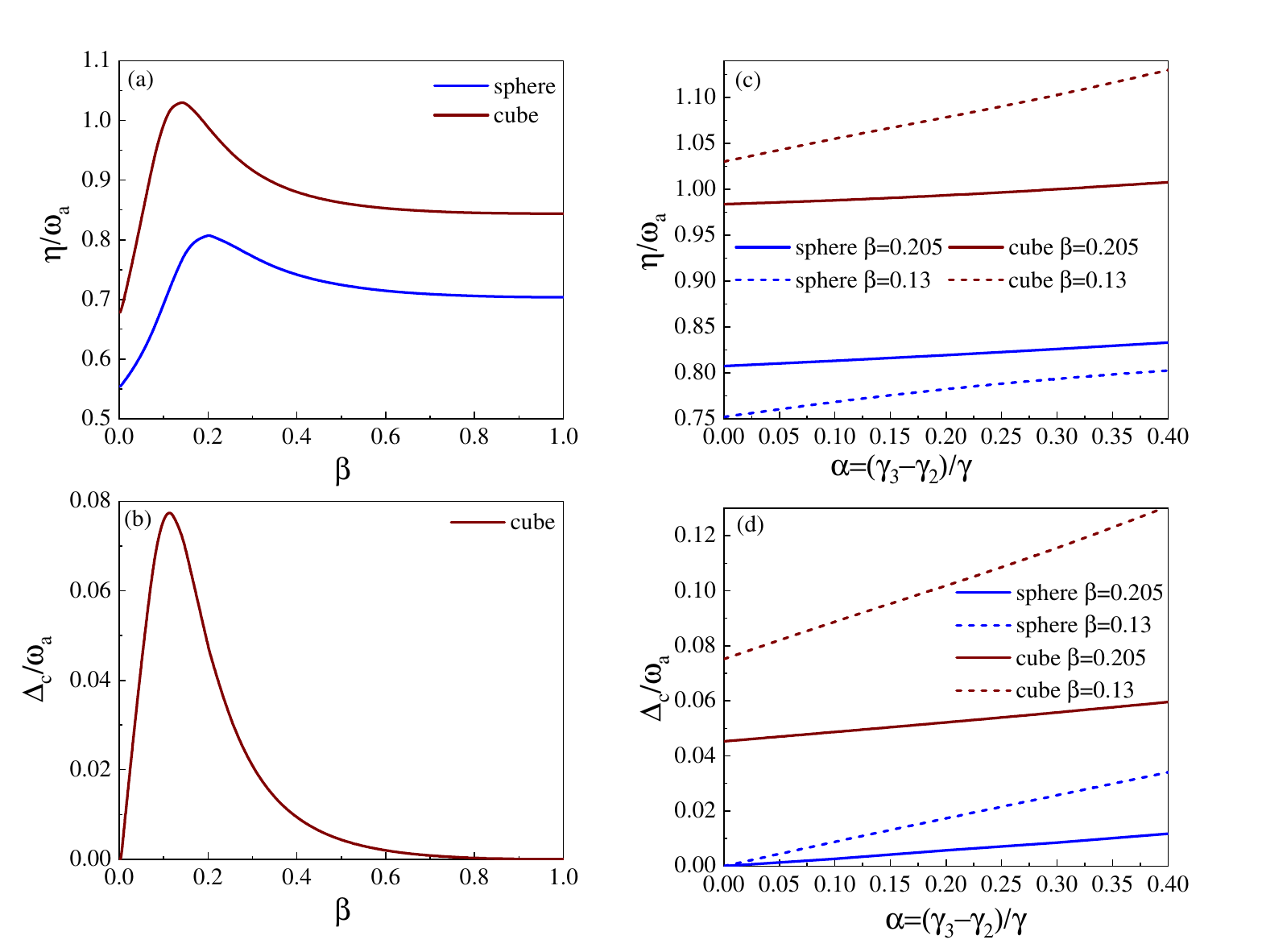}
    \caption{ Dimensionless isotropic $\eta/\omega_a$, (a,c), and   cubically-symmetric, $\Delta_c/\omega_a$, (b,d) electron-hole  exchange interaction   constants as functions of the light- to heavy hole mass ratio $\beta$ calculated  neglecting the valence band warping  for  spherical and  cube NCs (a,b), and  as functions of the valence band warping parameter $\alpha$ for value of light- to heavy hole mass ratio $\beta = 0.205 $    (CdSe parameters, solid lines) and $\beta = 0.13 $ (GaAs parameters, dashed lines) (c,d). } 
    \label{exch_calc_cube_beta}
\end{figure}

Firstly, we consider the holes in the spherical approximation for the Luttinger Hamiltonian. As  first shown in Ref. \cite{Efros1996},  the effective Hamiltonian for the  short-range exchange interaction  in spherical NCs can be written in the form of  Eq. \eqref{Hexch_sph}.  Here 
the parameter $\eta = \omega_{ a} \chi  $ includes only the  short-range contribution, and  $$\omega_{ a} = \frac{2}{3}\frac{w}{\pi a^3} $$ is the normalized exchange constant. 
The dimensionless parameter  $\chi$ depends only on the light to heavy hole mass ratio $\beta$ and can be found in Ref. \cite{Efros1996}.
The short-range part of parameter $\eta $ as a function of $\beta$ calculated for spherical NCs is shown in Fig. \ref{exch_calc_cube_beta} (a).

For a system with cubic symmetry inherited either from the cube shape of the NC or the valence band warping, the effective Hamiltonian for short-range electron-hole exchange interaction  can be written in the form of Eq. \eqref{Hcubic}.
We first consider the cubical NCs  with  the Luttinger Hamiltonian   in spherical approximation. In this case,   the cubically symmetric contribution to the exchange interaction is positive, and the ground exciton state is three-fold degenerate. The isotropic,  $\eta $,  and cubically-symmetric, $\Delta_c$, contributions to short-range exchange interaction for cube NCs as functions of $\beta$ are shown in Fig.  \ref{exch_calc_cube_beta} (a) and (b), respectively. One can see that the dependence of $\eta$ on $\beta$  looks very similar for spherical and cube NCs and $0<\Delta_c\ll \eta$.  

To illustrate the effect of the valence band warping in  Fig. \ref{exch_calc_cube_beta} (c) and (d) we show the isotropic,  $\eta $,  and cubically-symmetric, $\Delta_c$, contributions to the short-range exchange interaction as functions of the warping parameter $\alpha=(\gamma_3-\gamma_2)/\gamma$ calculated for $\beta=0.205$ corresponding to  CdSe and $\beta=0.13$ corresponding to GaAs. Here we consider only $\alpha>0$ because, to the best of our knowledge, in typical II-VI and III-V semiconductors $\gamma_3>\gamma_2$.  From the Figure, one can see the same effect  of the valence band warping for spherical and cube NCs: both $\eta$ and $\Delta_c$ increase with the increase of $\alpha$.      However, the ratio $0 < \Delta_c \ll \eta$ remains valid.

\subsection{In nanocrystals with  spheroidal and cuboidal shape}

\begin{figure}[h!]
    \centering
    \includegraphics[width=0.98\linewidth]{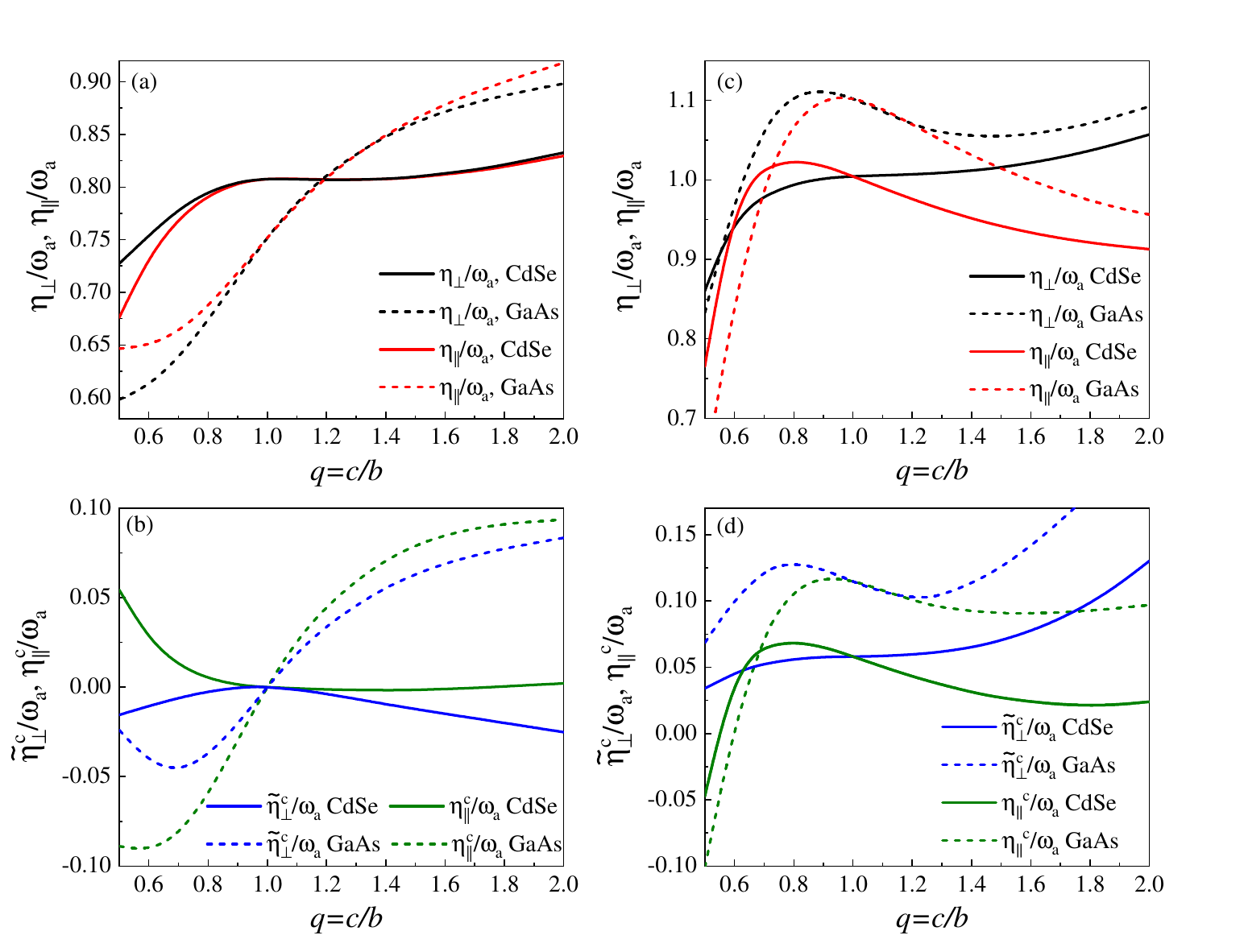}
    \caption{Dimensionless short-range exchange interaction constants (a) $\eta_{\perp}/\omega_a, \eta_{\parallel}/\omega_a$, and  (b) $ \tilde{\eta}_{\perp}^c/\omega_a, \eta_{\parallel}^c/\omega_a$ calculated for spheroidal NCs for  CdSe (solid lines) and  GaAs (dashed lines) parameters in spherical approximation for Luttinger Hamiltonian; Dimensionless short-range exchange interaction constants (c) $\eta_{\perp}/\omega_a, \eta_{\parallel}/\omega_a$ and (d) $\tilde{\eta}_{\perp}^c/\omega_a, \eta_{\parallel}^c/\omega_a $ calculated for cuboidal NCs with CdSe (solid lines) and GaAs (dashed lines) parameters with allowance of valence band warping.  } 
    \label{exch_calc_uniax}
\end{figure}

\begin{figure}[h!]
    \centering
    \includegraphics[width=0.99\linewidth]{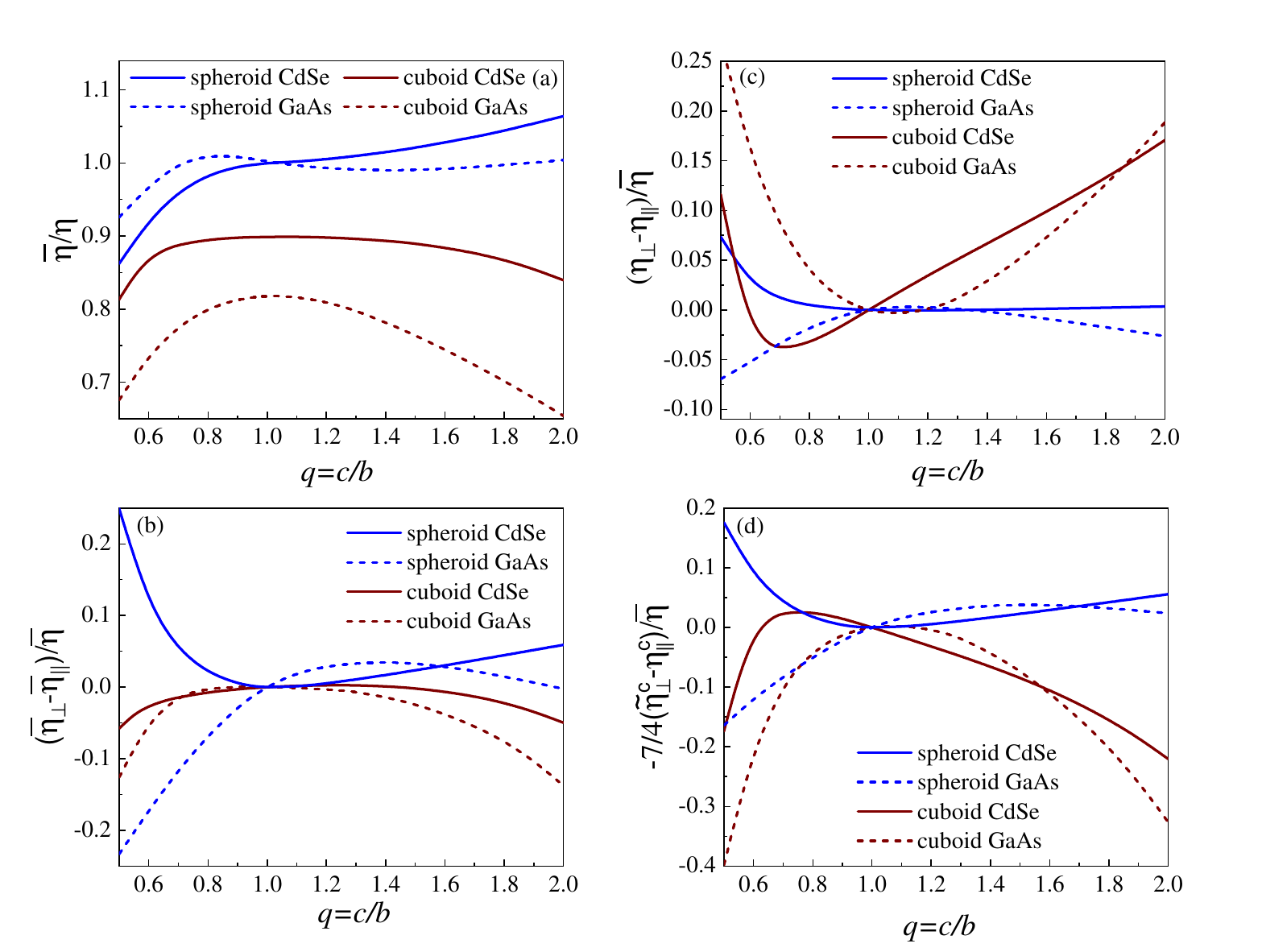}
    \caption{   (a) The  ratio $\bar{\eta}/\eta$ as the function of the   anisotropy parameter $q=c/b$  in spheroidal and cuboidal CdSe (solid lines) and GaAs (dashed lines) NCs; (b,c,d) The relative anisotropy of the short-range exchange constants  $(\bar{\eta}_\perp-\bar{\eta}_{\parallel})/\bar{\eta}$ (b),  $(\eta_\perp-\eta_{\parallel})/\bar{\eta}$ (c)  and  $-7/4(\tilde{\eta}^c_\perp-\eta^c_{\parallel})/\bar{\eta}$ (d) as function of the anisotropy parameter $q=c/b$.  }
    \label{exch_calc_os}
\end{figure}

\begin{figure}[h!]
    \centering
    \includegraphics[width=0.99\linewidth]{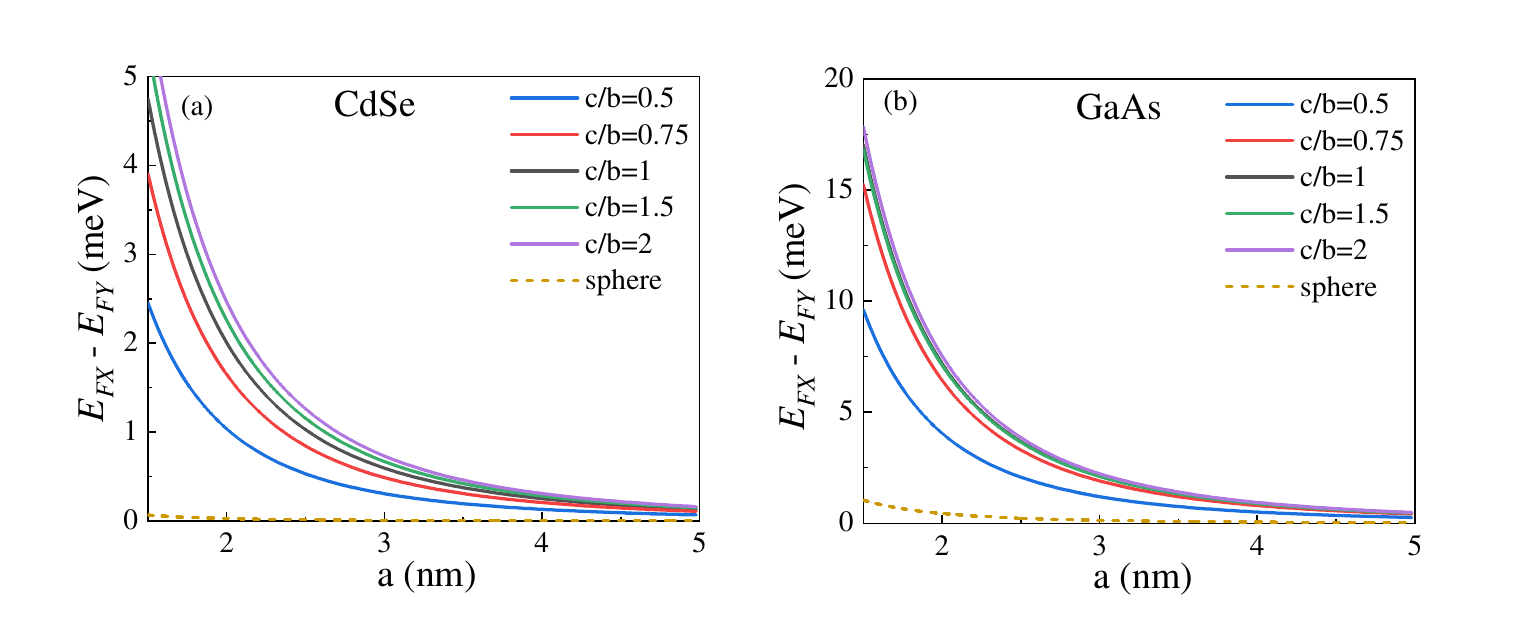}
    \caption{Estimations of dark exciton state with $|\mathcal F_z| =2$ splitting $E_{FX}-E_{FY}=3\eta_\perp^c$  in cuboidal  and spherical NCs with CdSe (a) and GaAs (b) parameters as function of unperturbed cube edge length and sphere radius $a$.   }
    \label{dark_calc_cube}
\end{figure}

The effective Hamiltonian of the short-range exchange interaction  for spheroidal NCs neglecting the cubic anisotropy of the crystal lattice, has the form shown in Eq. \eqref{uniax_exch}. The results of the calculation for spheroidal NCs with box-like infinite and parabolic potentials for CdSe band parameters in spherical approximation are shown in Fig. \ref{exch_calc_uniax} (a) and (b). Note that, unlike long-range exchange interaction, the contribution to the short-range interaction constants from the uniaxial anisotropy is non-zero only in the second order of perturbation theory with respect to the anisotropy parameter $\mu=c/b-1$. 
The effective Hamiltonian of the short-range exchange interaction for NCs with both cubic and uniaxial anisotropy has the form  given in Eq.  \eqref{cuboid}. The results of the calculation for cuboidal NCs with box-like infinite  potential for CdSe and GaAs band parameters, allowing for the valence band warping, are shown in Fig. \ref{exch_calc_uniax} (c) and (d). Our calculation has shown that the values of the linear exchange constants $\eta_{\perp}$ and $\eta_{\parallel}$ are roughly an order of magnitude larger than the values of the cubic constants $\tilde{\eta}_{\perp}^c$ and $\eta_{\parallel}^c$. However, the uniaxial anisotropy of the cubic exchange constants is comparable to the uniaxial anisotropy of the linear constants. We illustrate this effect below.  

We define the average exchange parameter $\bar{\eta}=(\tilde{\eta}_{\parallel}+2\tilde{\eta}_{\perp})/3$, where $\tilde{\eta}_{\parallel}$ and $\tilde{\eta}_{\perp}$ are introduced in Eq. \eqref{df}. Note that in the absence of uniaxial anisotropy $\bar{\eta}=\eta-7/4\Delta_c$, and in spherically-symmetric NCs $\bar{\eta}=\eta$.  Fig. \ref{exch_calc_os} (a)  shows the dependence of the short-range constants ratio $\bar \eta/\eta$ on the uniaxial anisotropy parameter $q=c/b$ calculated for CdSe (solid lines) and GaAs (dashed lines) NCs with spheroidal and cuboidal symmetry.  Here $\eta$ is the exchange interaction parameter in the absence of uniaxial anisotropy $q=c/b=1$, calculated for spheroidal and cuboidal CdSe and GaAs NCs.

Figs. \ref{exch_calc_os} (b,c,d) show the calculated relative anisotropy of the short-range exchange constants  $(\bar{\eta}_\perp-\bar{\eta}_{\parallel})/\bar{\eta}$ (b) and the partial contributions, $(\eta_\perp-\eta_{\parallel})/\bar{\eta}$ (c) and  $-7/4(\tilde{\eta}^c_\perp-\eta^c_{\parallel})/\bar{\eta}$ (d), respectively, as functions of the anisotropy parameter $q=c/b$ in spheroidal and cuboidal CdSe and GaAs NCs.   One can see that such anisotropy, although small, is not zero.  Moreover, as  mentioned above,   despite the order of magnitude difference in constant values, anisotropy for constants originating from the isotropic short-range exchange constant $\eta$ and the cubic constant $\Delta_c$  can be of the same order of magnitude. Moreover, taking into account the factor $7/4$, the anisotropy of cubic constants can be even larger. From the Figure, it can  also be seen that contributions to the value of $(\bar{\eta}_{\parallel}-\bar{\eta}_{\perp})/\bar{\eta}$ from $(\eta_\perp-\eta_{\parallel})/\bar{\eta}$ and  $(\tilde{\eta}^c_\perp-\eta^c_{\parallel})/\bar{\eta}$ can either add to each other (spheroidal NCs) or compensate (cuboidal NCs for most values of q).   

Although the cubically symmetric contributions to the electron-hole exchange interaction are quite small, they are responsible for the splitting of the $FX$ and $FY$ states.
This splitting $E_{FX}-E_{FY} = 3\eta_\perp^c$ is absent if only uniaxial anisotropy is considered without account  for the cubically symmetric contributions and $\pm 2$ dark exciton states remain degenerate. It is  small in the spherical NCs as it  comes only from the effect of the valence band warping. However,  it is noticeable in NCs with a cube shape and is additionally affected by the uniaxial anisotropy in cuboidal NCs as can be seen in Fig. \ref{dark_calc_cube}. One can see that the splitting is decreasing in oblate NCs with the increasing anisotropy and thus is expected to be small, for example, in typical quasi-two-dimensional nanoplatelets. Recently, the value $|E_{FX}-E_{FY}| \approx 20 \mu$eV was estimated in CdSe/CdS nanoplatelets  from the analysis of the experimentally observed effect of the dark exciton optical alignment \cite{Smirnova2023,Smirnova2025}.

\section{Discussion: exciton fine structure splitting and relative exciton oscillator strength. }\label{Discussion}

The exciton fine structure and the oscillator strengths of the exciton states depend substantially  on the parameters of the spin Hamiltonian, Eq. \eqref{spin}. Importantly, the exchange constants scale $\propto a^{-3}$ ($a$ is the NC size), while the anisotropic energy splitting $\Delta$ scales $\propto a^{-2}$ resulting in the dependence of the exciton fine structure on NC size. 
In addition to  the contributions  already considered , the  splitting of the excitons formed with heavy ($M=\pm 3/2$)  or light ($M=\pm 1/2$) holes may arise from the direct Coulomb interaction in NCs with uniaxial anisotropy.  In spherically or cubically-symmetric NCs with  a realized strong quantization regime for  both electrons and holes, the  Coulomb interaction does not split excitons with heavy and light holes.  In uniaxial NCs,  the Coulomb interaction does not split excitons with heavy and light holes in spheroidal NCs in the first order of perturbation theory; however, in higher orders of perturbation theory or in cuboidal NCs, the corresponding contribution would be non-zero, scaling as $\propto a^{-1}$. Our numerical estimations show that the Coulomb interaction contribution to parameter $\Delta$ for small NCs with the realized strong quantization regime is much smaller than the anisotropic splitting of the hole size quantization levels, scaling as $\propto a^{-2}$ and can be neglected.

\begin{figure}
\includegraphics[width=0.49\columnwidth]{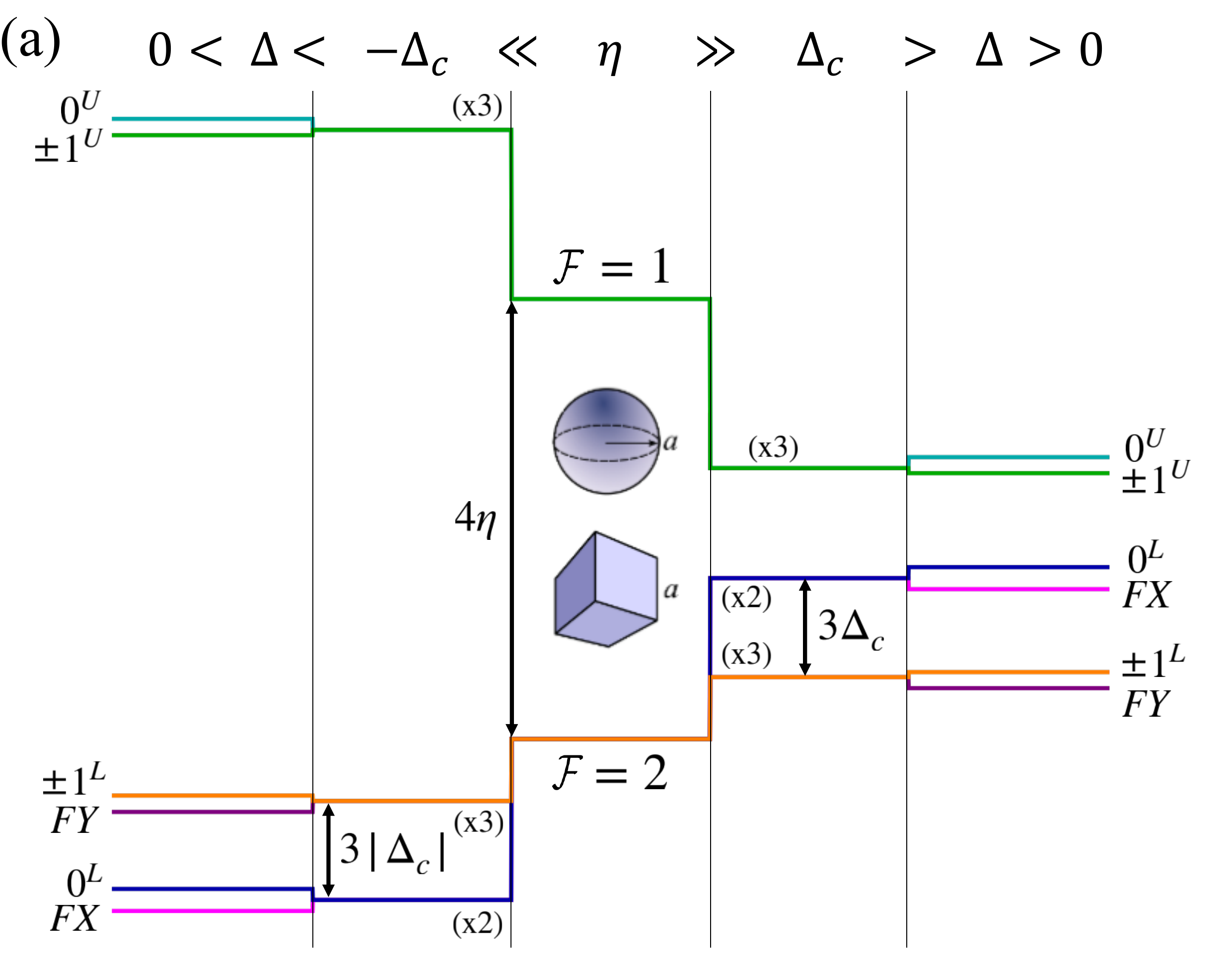}
\includegraphics[width=0.49\columnwidth]{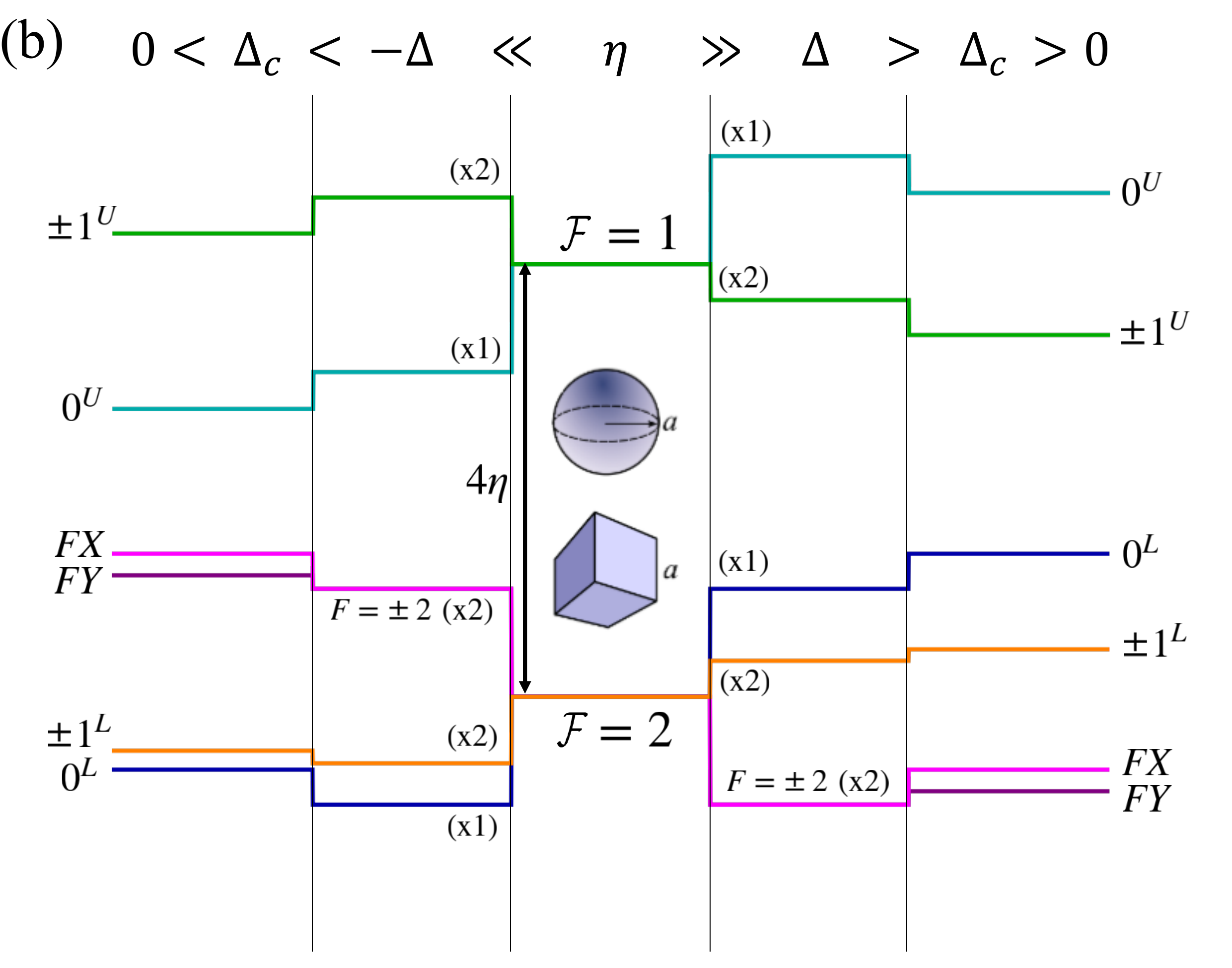}
   \caption{Schemes of the band edge exciton fine structure with the account of the isotropic electron-hole short-range exchange interaction ($\sim \eta$), the uniaxial anisotropy ($\sim \Delta$), and the cubic short-range exchange interaction ($\sim \Delta_c$); (a) $\eta \gg |\Delta_c|>\Delta>0$;   (b)  $\eta \gg |\Delta| >\Delta_c>0$. The middle part of each figure corresponds to the presence of the  isotropic exchange interaction only. } 
 \label{exch_fig}
\end{figure}

In Fig. \ref{exch_fig} we show  the sketches  of the exciton fine structure formed for different relations between constants (not to scale). The central (the third from the left) row on each panel corresponds to the largest contribution to the exciton fine structure, which is the isotropic part of the exchange interaction $4\eta>0$,  see Eq. \eqref{Hexch_sph}. Moving to the left/right, the other contributions are taken into account. The farther from the central row, the smaller the corresponding contribution becomes. In panel (a), the second largest contribution is the cubic exchange interaction $\sim \Delta_c$, positive to the right and negative to the left, the smallest contribution is from light and heavy hole splitting $\sim \Delta>0$.  In panel (b), the second largest contribution is  $\sim \Delta$, positive to the right and negative to the left, the smallest contribution is $\sim \Delta_c>0$.  One can see from  Fig. \ref{exch_fig} that the final ordering of the exciton sublevels might be quite different for different ratios between contributions. 

\begin{figure}[h!]
    \centering
    \includegraphics[width=0.99\linewidth]{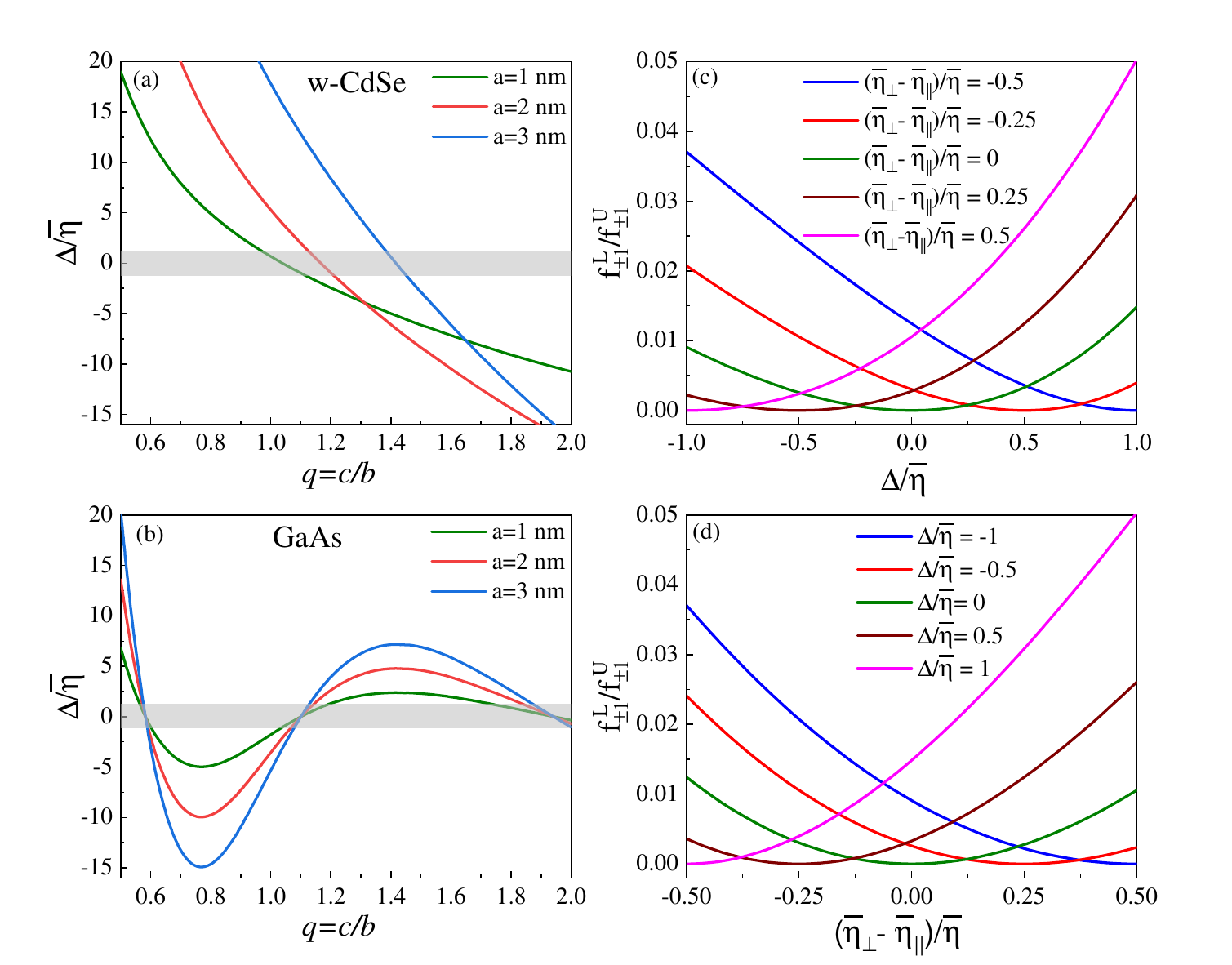}
    \caption{  The ratio $\Delta/\bar{\eta}$ calculated for  (a) w-CdSe and (b)  GaAs spheroidal NCs; (c) The relative oscillator strength of the $\pm 1^L$ exciton as  function  $\Delta/\bar{\eta}$ calculated for different the relative  anisotropy of the exchange constants $(\bar{\eta}_{\parallel}-\bar{\eta}_{\perp})/\bar{\eta}$; (d)  The relative oscillator strength of the $\pm 1^L$ exciton as of function $(\bar{\eta}_{\parallel}-\bar{\eta}_{\perp})/\bar{\eta}$  calculated for different values of $\Delta/\bar{\eta}$. } 
    \label{Delta_eta}
\end{figure}

In  small-sized  spheroidal or cuboidal  NCs with shapes close to spherical or cubic,  the largest are the linear exchange parameters $\eta_{\parallel}$ and $\eta_{\perp}$ originating from the isotropic parameter of the electron hole exchange interaction $\eta$. In the opposite case,  in large-sized NCs or in NCs with large uniaxial anisotropy, the  value of the anisotropic hole splitting $\Delta$ is expected to be the largest. The cubically-symmetric contributions to the exchange interaction as well as the uniaxial anisotropy of the exchange parameters are, as our calculations have shown, respectively small. It justifies the use of the simplified spin Hamiltonian in the form of Eq. (21) in  most cases, considering the electron-hole exchange interaction and the uniaxial anisotropy within first order perturbation theory. However,  there are several cases when    the anisotropic splitting $\Delta$ is small in comparison with the averaged exchange constant $\bar \eta$ even in the NCs with substantial  uniaxial shape anisotropy. The excitons in such nanocrystals with nearly degenerate hole states are often considered as isotropic degenerate excitons split by the isotropic electron-hole exchange interaction solely \cite{Brodu2019}. However, in such nanocrystals  an account of the uniaxial anisotropy of the exchange  constants, both linear and cubical, becomes important and may result into additional splitting of the exciton states. We consider two such cases as examples.  

The first case concerns the NCs with a wurtzite structure, where the splitting of the bulk valence band states by the crystal field is already present. In such NCs the additional anisotropy splitting  induced by the NC shape might compensate the effective crystal field splitting for holes $\Delta_{cr}=E_{1/2}-E_{3/2}$ and even reverse the level ordering \cite{Efros1996}. We are interested in the situation when $\Delta_{cr}$   compensates the anisotropic hole energy splitting $\Delta$ at $c/b\neq 1$. Such a situation for spheroidal w-CdSe NCs with a box-like potential  is illustrated in Fig. \ref{Delta_eta}(a). The compensation takes place in the prolate NCs with $q>1$.  The second case, the new one revealed in our calculations, is realized in spheroidal NCs made from a semiconductor with a small light to heavy hole mass ratio $\beta <0.14$, for example, GaAs.  In such semiconductors, the splitting in the case of the anisotropy axis directed along  the [001] axis has opposite signs at small and large aspect ratios, so that $\Delta$ is nonmonotonic function of $q=c/b$, see Fig. \ref{anis_en1} (d). As a result,  the ratio $\Delta/\bar \eta$ can be close to zero at quite large values of $|q-1|$ both in prolate and oblate NCs, see Fig. \ref{Delta_eta}(b). For the [111] direction of the anisotropy axis  the splitting remains monotonic function of $q$, however have much smaller absolute value than for [001] axis (see Fig. \ref{anis_en1} (d)). 
	
The calculations in Fig. \ref{Delta_eta} are done with the following parameters:  for w-CdSe $4\eta= 70/a[nm]^3$ meV, $\Delta_{cr}=25$ meV, $E_h=320/a[nm]^2$ meV \cite{Efros1996} and for  GaAs $4\eta=140/a[nm]^3$ meV, $E_h=670/a[nm]^2$ meV  \cite{Talapin2026}. It is worth to note that even in the zb-CdSe NCs a smaller but nonzero crystal field splitting may be present due to the presence of the stacking faults \cite{Golovatenko2022}.  For w-CdSe, we used our results for zb-CdSe as their band parameters are close, and for illustrative estimations, it is sufficient. We marked out by gray color in Figs. \ref{Delta_eta} (a) and (b) the area of relative anisotropy splitting $|\Delta/\bar \eta|< 0.5$.  For such NCs, the contributions to the exciton fine structure parameters  coming from the anisotropy of  exchange interaction constants may become important. We will focus the further discussion on such contributions.  

First, the uniaxial anisotropy affects the exciton fine energy splittings. In the absence of the uniaxial anisotropy, the  energy splitting between the dark, $\pm 1^{L}$, and bright, $\pm 1^U$, exciton states with projections $\pm 1$ (see Eq. \eqref{fine}) is given by  $\Delta E_{\pm 1}=4\bar{\eta}$. In the case of relatively small uniaxial anisotropy controlled by two parameters, $\Delta/\bar{\eta} \ll 1$ and $(\bar{\eta}_{\parallel}-\bar{\eta}_{\perp})/\bar{\eta} \ll 1$,  this splitting is described by $$\Delta E_{\pm 1} = 2\sqrt{f^2+d} \approx 4\bar \eta  + (\bar \eta_\perp - \bar \eta_\parallel)/3 - \Delta/2.$$ 
The uniaxial anisotropy mixes these states, making the lower $\pm 1^{L}$ exciton bright.
At the same time, the lowest exciton ground state for all parameters remains dark; its full momentum projection on the anisotropy axis, $F_z$, depends on the sign of the anisotropic parameters, as can be seen in Fig. \ref{exch_fig}.

As it is already mentioned, the  important effect of the uniaxial anisotropy  is the activation of the oscillator strength of the lower $\pm 1^L$ exciton originating from the dark ${\cal F}=2$ state, Eq. \eqref{ff}. Again, for the small anisotropy, it is controlled by  two dimensionless parameters $\Delta/\bar{\eta}$ and $(\bar{\eta}_{\parallel}-\bar{\eta}_{\perp})/\bar{\eta}$.  Fig.  \ref{Delta_eta} (c) shows the relative oscillator strength of the $\pm 1^L$ exciton as a  function  of $\Delta/\bar{\eta}$ for different  relative  anisotropy of the exchange constants $(\bar{\eta}_{\parallel}-\bar{\eta}_{\perp})/\bar{\eta}$.  Fig. \ref{Delta_eta} (d) the relative oscillator strength of the $\pm 1^L$ exciton as  function of $(\bar{\eta}_{\parallel}-\bar{\eta}_{\perp})/\bar{\eta}$ calculated for different values of $\Delta/\bar{\eta}$. From very similar Figs.  \ref{Delta_eta} (c) and (d) one can see that contributions to the oscillator strength of the $\pm 1^L$ exciton from anisotropic energy splitting $\Delta$ and anisotropy of the exchange interaction is comparable.  Note that they compensate each other at the point $\Delta=-2(\bar{\eta}_{\perp}-\bar{\eta}_{\parallel})$.  Note that in fully cubically-symmetric NCs, the activation of the lower $\pm 1^L$ exciton state would not happen, the presence of the uniaxial anisotropy is mandatory. For small uniaxial anisotropy, we obtain
$$\tilde f \approx  \sqrt{3} \left(1 - \frac{3}{4}\frac{\Delta+2(\bar{\eta}_{\perp}-\bar{\eta}_{\parallel})}{3\bar{\eta}+(\bar{\eta}_{\perp}-\bar{\eta}_{\parallel})}\right). $$
As a result, in  the range of parameters $\Delta/\bar{\eta}$ and $(\bar{\eta}_{\perp}-\bar{\eta}_{\parallel})/\bar{\eta}$ presented in Figs. \ref{Delta_eta} (c) and (d) the following expression
$$\frac{f_{\pm 1}^L}{f_{\pm 1}^U}\approx \frac{243(\Delta+2(\bar{\eta}_{\perp}-\bar{\eta}_{\parallel}))^2}{256(3\bar{\eta}+(\bar{\eta}_{\perp}-\bar{\eta}_{\parallel}))^4}$$
can be used as a good approximation. Note that the lowest exciton ${\pm 1}^L$ gets the oscillator strength not more than 5 \% from the upper ${\pm 1}^U$ bright exciton. Importantly, the lowest exciton states, $\pm 2$ or $\left|FX\right>$, $\left|FY\right>$, are not activated by the cubic symmetry or uniaxial anisotropic perturbations and remain dark unless the symmetry is lowered by the in-plane asymmetry mixing them with the upper $0^U$ bright exciton \cite{Goupalov2000} or breaking the time-inversion symmetry \cite{Sercel2017, Sercel2018}. For example,   the lowest dark exciton  ${\pm 2}$ might obtain the nonzero oscillator strength of the same order via the admixture of both ${\pm 1}^L$ and ${\pm 1}^U$ exciton states by the external magnetic field or internal exchange field \cite{Rodina2016}.

\section{Conclusion}\label{concl}

In this paper, we have considered the effects of cubical and uniaxial anisotropy originating from the crystal lattice symmetry and nanocrystal shape on the exciton fine structure, beyond the use of first order perturbation theory.  Our analysis reveals that,  in most practical cases, the effects of  uniaxial anisotropy on the electron-hole short range exchange interaction, including the appearance of cubic with respect to the hole momentum projections exchange constants, are small and can be neglected. 
However, an account of the cubically-symmetric contribution to the exchange interaction becomes important if the energy splitting of the dark $\pm 2$ exciton states is observed experimentally.
In addition, an account of the cubic symmetry of the crystal lattice is essential for the correct description of the anisotropic splitting of the hole states with full momentum projections $\pm 3/2$ and $\pm 1/2$ on the anisotropy axis of the nanocrystal, depending on its orientation with respect to the crystallographic directions. 

An important special case arises when the hole anisotropic energy splitting becomes small in substantially uniaxial structures. In this case, the uniaxial anisotropy of the electron-hole exchange interaction parameters, both linear and cubic with respect to the hole momentum projections, should be considered. It affects the fine structure splitting and oscillator strength of exciton states and might either  compensate for or enhance the effect of the hole anisotropic splitting.  We emphasize that the hole uniaxial-anisotropic splitting depends strongly on the direction of the anisotropy axis with respect to the crystallographic axes, especially for semiconductors with a light-to-heavy hole mass ratio $\beta$ close to $0.14$. Thus, an accurate analysis of the exciton fine structure, for example, in recently reported GaAs spheroidal NCs  \cite{Talapin2024,Talapin2026} should account for both the crystal lattice symmetry and the anisotropy axis direction. 

The symmetry consideration and the numerical calculations presented here for the $D_{2d}$ point symmetry group  can be generalized  to the $C_{2v}$ point symmetry group, which describes nanocrystals with additional anisotropy in the plane perpendicular to the main axis. Such an extension would allow one to quantitatively compare the contributions to the anisotropic splitting of $\pm 1^L$ excitons  coming from the in-plane anisotropy of the long-range and the short-range cubic exchange interaction,  thereby allowing a direct test of the recent hypothesis regarding their mutual compensation in CdSe/CdS nanoplatelets \cite{Smirnova2025}.    

\begin{acknowledgments}
The authors thank E.L. Ivchenko and I.D. Avdeev for valuable discussions. The work was supported  by the  Russian Science Foundation under project No. 23-12-00300-$\Pi$. 
	\end{acknowledgments}

\appendix

\setcounter{equation}{0}
\setcounter{figure}{0}
\setcounter{table}{0}
\renewcommand{\thefigure}{A\arabic{figure}}
\renewcommand{\theequation}{A\arabic{equation}}
\renewcommand{\thetable}{A\arabic{table}}

\section{The hole kinetic energy Hamiltonian after the coordinate replacement}\label{AA}

The explicit form of the correction to the hole Hamiltonian \eqref{lutt} for anisotropy axis $z'||z||[001]$:
\begin{equation}\label{lutt_matrix_an}
\hat{H}_{an}^{001}=\frac{\hbar^2}{2m_0}\left(
\begin{array}{cccc}
 P+Q & H & I &
   0 \\
 H^\dag &  P-Q & 0 & I \\
 I^\dag & 0 &  P-Q & -H \\
 0 & I^\dag & -H^\dag &  P+Q \\
\end{array}
\right),\end{equation}
$$P=\gamma_1\left[\left(q^{2/3}-1\right) \left(k_x^2+k_y^2\right)+k_z^2
   \left(q^{-4/3}-1\right)\right],$$
$$Q=\gamma_2\left[\left(q^{2/3}-1\right) \left(k_x^2+k_y^2\right)-2k_z^2
   \left(q^{-4/3}-1\right)\right],$$
$$H=-2 \sqrt{3} \gamma_3 (q^{-1/3}-1)k_z (k_x-\mathrm i k_y),$$  $$I=-\sqrt{3} (q^{2/3}-1)(\gamma_2 k_x^2- \gamma_2 k_y^2-2\mathrm i\gamma_3k_xk_y).$$

The correction to the hole Hamiltonian for anisotropy axis $z'||z||[001]$ and arbitrary anisotropy $q=c/b$
\begin{equation}\label{lutt_matrix_an111}
\hat{H}_{an}^{111}=\frac{\hbar^2}{2m_0}\left(
\begin{array}{cccc}
 P^{111}+Q^{111} & H^{111} & I^{111} &
   0 \\
 H^{111\dag } &  P^{111}-Q^{111} & 0 & I^{111} \\
 I^{111\dag } & 0 &  P^{111}-Q^{111} & -H^{111} \\
 0 & I^{111\dag } & -H^{111\dag } &  P^{111}+Q^{111} \\
\end{array}
\right),\end{equation}
$$P^{111}=\gamma_1\left[(k_x^2+k_y^2+k_z^2)\left(\frac{2}{3}q^{\frac{2}{3}}+\frac{1}{3}q^{-\frac{4}{3}}-1\right)-\frac{2}{3}(k_xk_y+k_yk_z+k_xk_z)\left( q^{\frac{2}{3}}-q^{-\frac{4}{3}}\right)\right],$$
$$Q^{111}=\gamma_2\left[(k_x^2+k_y^2-2k_z^2)\left(\frac{2}{3}q^{-\frac{1}{3}}+\frac{1}{3}q^{\frac{2}{3}}-1\right)-\frac{2}{3}\left(2k_xk_y-k_yk_z-k_xk_z\right)\left( q^{\frac{2}{3}}-q^{-\frac{1}{3}}\right)\right],$$
$$H^{111}=-2 \sqrt{3} \gamma_3\left[ k_z(k_x-\mathrm{i}k_y)\left(\frac{2}{9}q^{-\frac{4}{3}}+\frac{2}{9}q^{-\frac{1}{3}}-\frac{5}{9}q^{\frac{2}{3}}-1\right)\right]-$$
$$-2 \sqrt{3} \gamma_3\left[\frac{1}{9}(k_x^2+k_y^2+k_z^2)(1-\mathrm{i})\left(q^{-\frac{4}{3}}+q^{-\frac{1}{3}}-2q^{\frac{2}{3}}\right)+\frac{1}{3}(k_y^2-\mathrm{i}k_x^2)\left(-q^{-\frac{1}{3}}+q^{\frac{2}{3}}\right)\right]-$$
$$-2 \sqrt{3} \gamma_3\left[-\frac{1}{9}\left(k_xk_y(1-\mathrm{i})+k_z(k_y-\mathrm{i}k_x)\right)\left(2q^{\frac{4}{3}}-q^{-\frac{1}{3}}-q^{\frac{2}{3}}\right)\right]$$

$$I^{111}=-\sqrt{3}\gamma_2\left[(k_x^2-k_y^2)\left(\frac{2}{3}q^{-\frac{1}{3}}+\frac{1}{3}q^{\frac{2}{3}}-1\right)-\frac{2}{3}k_z(k_x-k_y)\left(q^{\frac{2}{3}}-q^{-\frac{1}{3}}\right)\right]+$$
$$+4\mathrm{i}\sqrt{3}\gamma_3\left[k_xk_y\left(\frac{2}{9}q^{-\frac{4}{3}}+\frac{2}{9}q^{-\frac{1}{3}}+\frac{5}{9}q^{\frac{2}{3}}-1\right)-\frac{1}{9}k_z^2\left(q^{-\frac{4}{3}}-2q^{-\frac{1}{3}}+q^{\frac{2}{3}}\right)\right]+$$
$$+4\mathrm{i}\sqrt{3}\gamma_3\left[\frac{1}{9}k_z(k_x+k_y)\left(-2q^{-\frac{4}{3}}+q^{-\frac{1}{3}}+q^{\frac{2}{3}}\right)-\frac{1}{9}(k_x^2+k_y^2)\left(q^{-\frac{4}{3}}+q^{-\frac{1}{3}}-2q^{\frac{2}{3}}\right)\right].$$

The calculations with anisotropic Hamiltonian are done numerically using the  numerical method introduced in Refs. \cite{Semina2015, Semina2016} for spheroidal NCs with parabolic (or Gaussian) potential and modified in Refs. \cite{Semina2021,Semina2023} for spheroidal and cuboidal NCs with box-like infinite potential.  As shown in Ref. \cite{Semina2016} for spheroidal NCs with parabolic (or Gaussian) potential, the numerical calculations with the anisotropic kinetic energy Hamiltonian obtained after the   coordinate replacement and isotropic potential energy give the same results as the direct numerical calculation with initial $\hat H_L$ and anisotropic NC potential.  In this work, we also checked this for cuboidal NCs with box-like infinite potential. For  spheroidal   NCs with box-like potential, the numerical method is realized for the anisotropic kinetic energy Hamiltonian only.

\setcounter{equation}{0}
\setcounter{figure}{0}
\setcounter{table}{0}
\renewcommand{\thefigure}{C\arabic{figure}}
\renewcommand{\theequation}{C\arabic{equation}}
\renewcommand{\thetable}{C\arabic{table}}

\setcounter{equation}{0}
\setcounter{figure}{0}
\setcounter{table}{0}
\renewcommand{\thefigure}{D\arabic{figure}}
\renewcommand{\theequation}{D\arabic{equation}}
\renewcommand{\thetable}{D\arabic{table}}


\end{document}